# Scalability of Superconductor Electronics: Limitations Imposed by AC Clock and Flux Bias Transformers

Sergey K. Tolpygo, *Senior Member, IEEE*

*Abstract*—**Flux transformers are the necessary component of all superconductor digital integrated circuits utilizing ac power for logic cell excitation and clocking, and flux biasing, e.g., Adiabatic Quantum Flux Parametron (AQFP), Reciprocal Quantum Logic (RQL), superconducting sensor arrays, qubits, etc. On average, one transformer is required per one Josephson junction. We consider limitations to the integration scale (device number density) imposed by the critical current of the ac power transmission lines (primary of the transformers) and cross coupling between the adjacent transformers. The former sets the minimum linewidth and the mutual coupling length in the transformer, whereas the latter sets the minimum spacing between the transformers. Decreasing linewidth of superconducting (Nb) wires increases kinetic inductance of the transformer's secondary, decreasing its length and mutual coupling to the primary. This limits the minimum size. As a result, there is a minimum linewidth $w_{min} \sim 100$ nm which determines the maximum achievable scale of integration. Using AQFP circuits as an example, we calculate dependences of the AQFP number density on linewidth for various types of transformers and inductors available in the SFQ5ee fabrication process developed at MIT Lincoln Laboratory and estimate the maximum circuit density as a few million AQFPs per cm$^2$. We propose an advanced fabrication process for a 10x increase in the density of AQFP and other ac-powered circuits. In this process, inductors are formed from a patterned bilayer of a geometrical inductance material (Nb) deposited over a layer of high kinetic inductance material (e.g., NbN). Individual pattering of the bilayer layers allows to create stripline inductors in a wide range of inductances, from the low values typical to Nb striplines to the high values typical for NbN thin films, and preserve sufficient mutual coupling in stripline transformers with extremely low crosstalk.**

## I. INTRODUCTION

SUPERCONDUCTOR digital electronics easily beats CMOS and prospective beyond CMOS technologies in such important performance metrics as energy dissipation and processing speed. Superconductor single flux quantum (SFQ) electronics [1] holds the record in clock speed of simple circuits of about 770 GHz [2] and in its slower, adiabatic implementations can operate with energy per bit near the Landauer's thermodynamic limit $k_B T \ln 2$ [3]-[6]. However, these performance advantages so far have not benefited any large-scale computational system because integration scale of superconductor digital circuits is currently three to four orders of magnitude lower than of the CMOS circuits. For instance, the largest demonstrated circuits in superconductor electronics have about one million Josephson junctions [7], [8] whereas the modern CMOS circuits have over 50 billion transistors, a 50,000× difference [9].

Due to the recent progress in fabrication technology of niobium-based superconductor integrated circuits, the minimum feature size was reduced to 120 nm [10], [11]. This allowed for an increase in the circuit density to about $1.5 \cdot 10^7$ Nb/Al-AlO$_x$/Nb Josephson junctions (JJs) per square centimeter [10], [12], about ten-fold increase from the previous level [7]. For a comparison, the present density of CMOS circuits is 1000x higher, about $1.4 \cdot 10^{10}$ transistors per cm$^{-2}$ [9]. The largest demonstrated density of superconductor Josephson junction-based random access memory (JRAM) is 1 Mbit cm$^{-2}$ [13]. For a comparison, RAM technology based on spin-transfer torque magnetic RAM, the so-called STT-MRAM, operating at room temperature has a 1000x higher density [14], although it uses devices similar to multilayered sandwich-type Josephson junctions.

In [15], the author argued that superconductor digital electronics is fundamentally less scalable than semiconductor electronics because of the fundamental difference in information encoding. Indeed, in superconductor electronics information is encoded, stored, and transferred by magnetic flux quanta created by superconducting currents circulating in

This material is based upon work supported by the Under Secretary of Defense for Research and Engineering under Air Force Contract No. FA8702-15-D-0001.

Sergey K. Tolpygo is with Lincoln Laboratory, Massachusetts Institute of Technology, Lexington, MA 02421 (e-mail: sergey.tolpygo@ll.mit.edu).

closed superconducting loops (inductors) interrupted by Josephson junctions. In semiconductor electronics, information is encoded by a stationary electric charge on the gates of field-effect transistors. Obviously, localized charge on a capacitor occupies less space than the moving charge – superconducting loop current. Hence, charge-based devices can always be made smaller and their circuits made denser and more scaled-up than flux-based devices and circuits.

The goal of this work is to establish fundamental limits on the scalability, i.e., the maximum circuit density, of ac-powered superconductor digital electronics, imposed by two basic components of all superconductor integrated circuits: inductors and transformers. Limitations imposed by sandwich-type (trilayer) Josephson junctions like Nb/AlO$_x$-Al/Nb were discussed in [15].

A problem of superconducting transformers has emerged with advancement of ac powered and ac clocked superconductor logic solutions instead of the dc-powered RSFQ logic [1]. Starting from the original Parametric Quantron (PQ) [3], [16], [17] and going to its analogs, renames and incarnations − DC Flux Parametron [18], [19], Quantum Flux Parametron (QFP) [20], [21], and Adiabatic Quantum Flux Parametron (AQFP) [22], [23] − all logic solutions based on parametric devices require a multiphase ac excitation. These ac signals are inductively coupled (via transformers) to the devices in order to modulate their Josephson inductance (critical current of Josephson junctions) and produce a change in the logic state and a parametrically amplified output current upon applying a weak input current $I_{in}$. The only exception is a dc-powered nSQUID logic whose devices (nSQUIDs) use transformers to create a large negative mutual inductance between the SQUID arms [4], [24].

Another example is Reciprocal Quantum Logic [25], [26] which requires four-phase ac power delivered via transformers to propagate positive and negative single flux quantum (SFQ) pulses in the same direction along Josephson transmission lines (JTLs) and provide energy to and synchronization of RQL gates.

In addition to the power and clock delivery, in all types of superconductor logic and memory circuits, superconducting qubit circuits, superconducting sensor arrays, etc. superconducting flux transformers (mutual inductors) are used to provide flux biasing, signal inverting (NOT function), and coupling between cells.

As an example, we will consider what is now known as AQFP cell shown in Fig. 1a. It is identical to QFP and PQ cells, and may only slightly differ in parameter values. The AQFP logic requires, at least, one ac transformer per Josephson junction. Therefore, the circuit density (device number density) cannot be higher than the density of the transformers. A very similar analysis to the offered below can be easily done for RQL gates and their JTLs, which typically require one transformer per two junctions [25]-[27], and for any other ac-powered logic, memory, and quantum circuits.

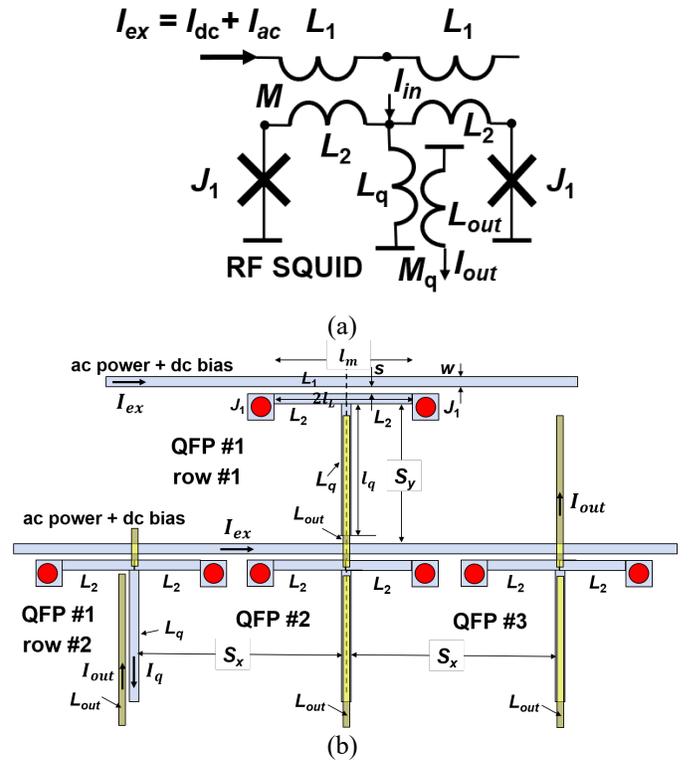

(a)

(b)

**Fig. 1.** **(a)** Schematics of PQ, QFP and AQFP cells. It consists of two identical RF SQUIDs connected in parallel. In the adiabatic regime of operation, the typical parameters of AQFPs are $\beta_L \equiv 2\pi I_{c1}L_2/\Phi_0 = 0.2$, $\beta_{Lq} \equiv 2\pi I_{c1}L_q/\Phi_0 = 1.6$, and $I_{c1} = 50\ \mu$A is the critical current of junctions $J_1$ [34].

**(b)** A sketch of a planar transformer between the primary inductor $L_1$, a part of the ac-power transmission line, and the QFP cell inductors $L_2$ which are either a microstrip (one ground plane) or stripline (two ground planes) inductors with length $l_L$, laying in the same or adjacent plane as the $L_1$. The mutual running length of the inductors $L_1$ and $L_2$, which determines their total mutual inductance $M$, is $l_m$. The full QFP consists of two connected half-cells (RF SQUIDs). Also shown is the second row of QFPs. Their inputs are inductively coupled to other QFPs via the output inductor $I_{out}$. Spacing $S_y$ between the QFP rows (or the ac power lines) determines their cross coupling – ac excitation in QFP#2 produced by the power line of the QFP#1 in the same column, and vice versa. Spacing $S_x$ determines cross coupling between the output transformers $L_q$, $I_{out}$ of QFPs in the same row. For layout examples, see [37]. Inductor $L_q$ needs to be place perpendicular to inductors $L_1$ and $L_2$ in order to minimize direct coupling of the ac excitation to the output.

## II. AC Excitation and DC Flux Transformers: Physical Limitations

### A. Main Features of ac Excitation Transformers

The typical use of transformers in superconductor electronics is to provide a dc flux bias and/or ac flux excitation to logic cells with amplitude $\frac{\Phi_0}{\gamma}$; typically, $\gamma = 2$. Very often, e.g., in AQFP, both dc flux bias $\Phi_{dc} = \Phi_0/2$ and ac excitation are required. They can be applied either via the same or two separate transformers. For simplicity, we consider only the former case, requiring $\gamma = 1$, because using two transformers increases the total transformer area by about 2x.

In order to provide the required dc bias and ac excitation, the mutual running length of the transformer primary wire and the transformer secondary wire should be

$$l_m = \Phi_0/(\gamma M_l I_{ex}), \tag{1}$$

where $I_{ex} = I_{dc} + I_{ac}$ is the amplitude of the total excitation current that is fed into the $L_1$, and $M_l$ is the mutual inductance per unit length between the wires of the primary and secondary.

On the other hand, the transformer secondary, $L_2$ is always a part of the logic cell and its inductance is determined by the cell design parameter

$$\beta_L = 2\pi I_{cJ} L_2/\Phi_0, \tag{2}$$

where $I_{cJ}$ is the Josephson junction critical current of junction $J_1$. Hence, length of the inductor $L_2$ is given by

$$l_L = \frac{\beta_L \Phi_0}{2\pi L_{Ll} I_{cJ}}, \tag{3}$$

where $L_{Ll}$ is the inductance per unit length of the secondary.

Obviously, the transformer with the smallest area in this simple two-wire configuration can only be formed if $l_m \leq l_L$ or if

$$I_{ex} \geq \frac{2\pi l_L I_{cJ}}{\gamma \beta_L M_l}. \tag{4}$$

Multiturn transformers with $l_m > l_L$ are possible but will not be considered here because they always are occupy much larger area and, hence, restrict the circuit density more than the simplest parallel-wire transformer shown in Fig. 1b.

The mutual inductance of any two conductors is smaller than the magnetic part of the self-inductance $L_M$ of the conductors

$$M = \kappa (L_{M1} L_{M2})^{1/2}, \tag{5}$$

$$L_l = L_{Ml} + L_{Kl}, \tag{6}$$

$$L_{Kl} = \mu_0 \frac{\lambda^2}{tw}, \tag{7}$$

where $L_{Kl}$ is kinetic inductance per unit length of a wire with thickness $t$ and width $w$, $\lambda$ is the magnetic field penetration depth; $\kappa \leq 1$ is the coupling coefficient between the magnetic inductances of the primary, $L_{M1}$ and the secondary, $L_{M2}$. If the primary and secondary are formed by inductors of the same type with equal magnetic inductance per unit length, $L_{M1l} = L_{M2l} = L_{Ml}$, the required primary excitation current from (4) is

$$I_{ex} \geq \frac{2\pi}{\gamma \beta_L \kappa} \left(1 + \frac{\mu_0 \lambda^2}{twL_{Ml}}\right) I_{cJ}. \tag{8}$$

This current is significantly larger than the $I_{cJ}$ and *must increase with decreasing the cross section of the inductors*, i.e., with increasing the scale of integration.

On the other hand, $I_{ex}$ must be smaller than some maximum current $I_{max}$ related to the critical current, $I_c$ of superconducting (Nb) wires forming the transformer,

$$I_{ex} \leq I_{max} = \alpha I_c = \alpha j_c wt, \tag{9}$$

where $\alpha < 1$ is a safety factor and $j_c$ is the superconductor critical current density. Hence, the minimum possible inductor cross sectional area $A = wt$ can be found from the condition

$$wt \geq \frac{2\pi}{\alpha \beta_L \gamma j_c \kappa} \left(1 + \frac{\mu_0 \lambda^2}{twL_{Ml}}\right) I_{cJ}, \tag{10}$$

which is not a quadratic equation because of a nontrivial dependence of $L_{Ml}$ and $\kappa$ on $t$ and $w$.

In superconductors, the fundamental limit to the $j_c$ is the Ginzburg-Landau critical current density $j_c^{GL}$ given by

$$j_c^{GL} = \frac{2\sqrt{2}}{3\sqrt{3}} \frac{B_c}{\mu_0 \lambda}, \tag{11}$$

where $B_c$ is the thermodynamic critical magnetic field of the superconductor and $\lambda$ is the magnetic field penetration depth [28]. Using the microscopic theory [29], Bardeen expressed this depairing critical current density as

$$j_c^{GL} \approx \frac{B_{c0}}{\mu_0} \left(\frac{\Delta_0}{\hbar \rho}\right)^{\frac{1}{2}} (1 - \frac{T^2}{T_c^2})^{\frac{3}{2}}, \tag{12}$$

where $B_{c0}$ and $\Delta_0$ are the thermodynamic critical field and the energy gap at zero temperature $T = 0$, respectively; $\rho$ is the film resistivity in the normal state [30].

For niobium, $B_c(T) = B_{c0}(1 - T^2/T_c^2)$, $T_c = 9.1$ K and $B_{c0} = 0.155$ T, giving $B_c = 0.122$ T at 4.2 K. The measured value of the penetration depth in our Nb films is $\lambda = 90$ nm. Therefore, $j_c^{GL}(4.2\text{ K}) = 0.59$ A/$\mu$m$^2$ for our Nb films at 4.2 K. The measured critical current density $j_c$ in Nb wires with $w \approx t \approx 200$ nm is a factor of two lower than this theoretical value, about 0.37 A/$\mu$m$^2$ [10], [31]. Since $j_c$ may decrease further with reducing the wire dimensions, e.g., due to increasing Nb contamination and resistivity, it would be practical to take $\alpha j_c = 0.25$ A/$\mu$m$^2$, a 33% lower value than the measured critical current density, in order to have some safety margin.

To proceed further with (10) and find the dependence of the transformer area on the linewidth, we need to specify the type of the transformer in order to determine $L_{Ml}$ and $\kappa$. Before going to numerical results, as a simple illustration, we consider a transformer formed by two parallel microstrips laying in the same plane.

For superconducting microstrips with rectangular cross section, magnetic part of the inductance per unit length is given in [32] by

$$L_{Ml} = \frac{\mu \mu_0}{4\pi} \ln\left[1 + \frac{4(d_1 + \frac{t}{2} + \lambda)^2}{0.2235^2(w+t)^2}\right], \tag{13}$$

where $d_1$ is the dielectric thickness between the microstrip and the ground plane, and $\mu$ is magnetic permeability of the dielectric, hereafter assumed to be 1. From the fabrication process practicality, we are interested in wires with nearly square cross sections $w \approx t$ and deeply scaled features with $w, t \ll 2(d_1 + \lambda)$. In this case, the minimum cross section area is achieved at the maximum possible coupling $\kappa = 1$ in (10) and given by solution of the equation

$$A \geq \frac{2\pi}{\alpha \beta_L \gamma j_c} \left(1 + \frac{4\pi \lambda^2}{A\mu \ln(1 + \frac{(d_1 + \lambda + t/2)^2}{0.05A})}\right) I_c. \tag{14}$$

To solve (14), let us use the typical parameters of ac excitation and flux bias transformers in AQFPs as an example: $I_c = 50$ $\mu$A, $\gamma = 1$, $\beta_L \approx 0.2$ [22], [33]. In the widely used fabrication processes SFQ5ee [34], transformers using Nb microstrips M6aM4 with $d_1 = 615$ nm (signal trace on the layer M6 above the ground plane M4) are the most convenient because they are the closest to the layer of Josephson junctions. Then, solving (14) gives the minimum cross section area $A_{min} = 1.34 \cdot 10^{-2}$ μm². Hence, the typical minimum linewidth and the film thickness are $(A_{min})^{\frac{1}{2}} \approx 116$ nm. *At smaller cross sections, the required transformer cannot be made because the excitation current required to induce $\Phi_0$ flux amplitude in the transformer secondary would exceed the critical current of the transformer primary wire.* In reality, the coupling coefficient is always less than 1. For a more realistic $\kappa = 0.5$, $A_{min} = 2.18 \cdot 10^{-2}$ μm² and $(A_{min})^{\frac{1}{2}} = 148$ nm.

The minimum linewidth following from the minimum cross section area) determines the minimum possible coupling length between the transformer primary and secondary, i.e., the minimum possible size of the ac-powered cell along the ac power transmission line. Existence of the minimum linewidth and the minimum cell size is the first limit on scalability of ac-powered superconductor electronics (AQFP, RQL, etc.) caused by the finite superconducting critical current of the primary wire in the ac and dc flux-biasing transformers. This limit can be reached in a 90-nm technology node (90-nm linewidth). Further reduction of the linewidth would not significantly increase the density of superconductor integrated circuits using ac powering of logic gates. This is our first conclusion.

In the following sections we will consider other limitations on the cell height (in the *y*-direction perpendicular to the ac power transmission line) and width. For specificity, we will use the typical parameters of AQFPs, whereas the same arguments and estimates apply to other types of ac-powered superconductor logics.

### B. Inductor $L_q$ and the Output Coupling Transformer

The length of the inductor $L_q$ in AQFP, $l_q$ is given by

$$l_q = \frac{L_q}{L_{ql}} = \frac{\beta_{Lq}\Phi_0}{2\pi L_{ql}l_{cJ}}, \tag{15}$$

Here $L_{ql}$ is inductance per unit length of inductor $L_q$ which may differ from the per length inductance of the excitation transformer secondary $L_{Ll}$. The typical AQFP parameters are $I_{cJ} = 50$ μA and $\beta_{Lq} = 1.6$, giving $L_q = 10.53$ pH. At $L_{ql} \sim 1$ pH/μm, $l_q$ is about 10 μm.

Inductor $L_q$ *needs to be placed perpendicular to the inductors $L_1$ and $L_2$, forming a T-shape, in order to minimize direct coupling of the ac excitation to the output.* The aspect ratio of this "T" in the typical AQFP cell (width to length ratio) is $2\frac{l_L}{l_q} = 1:4$ because the ratio of the optimal parameters $\frac{\beta_q}{\beta_L} = \frac{l_q}{l_L}$ is 1.6:0.2=8:1. The minimum possible area of the typical AQFP is $A_{QFP} = 2l_L l_q = \frac{\beta_L \beta_q \Phi_0^2}{2\pi^2 L_L L_{ql} l_{cJ}^2}$ if the area occupied by the junctions $J_1$, the transformer primary, and vias between the inductors can be neglected. The AQFP maximum number density is then

$$n_{QFP} = \chi A_{QFP}^{-1} = \frac{2\pi^2 \chi L_{Ll}L_{ql}l_{cJ}^2}{\beta_L \beta_q \Phi_0^2}, \tag{16}$$

where $\chi < 1$ is the area filling factor.

Using the AQFP parameters in Fig. 1 and the maximum expected value of magnetic inductances $L_{Ll} = L_{ql} \sim 1$ pH/μm, we get $n_{QFP} < 3.6 \cdot 10^6$ cm⁻². This is our second conclusion: *the number density of AQFPs using Nb inductors is limited from above by three to four million AQFPs per cm²*, corresponding to about 1M cm⁻² AQFP majority gates composed of three (MAJ3) AQFPs.

This estimate does not account for possible limitations imposed by the supercurrent-carrying capacity of the ac transformer and for a possibility of using inductors with higher $L_{ql}$ values. Accounting for these factors does not change the order of magnitude in the AQFP density estimate, but requires a more detailed analysis.

The superconducting material and cross section of wires for inductors $L_q$ and $L_{out}$ are selected such that both currents $I_q$ (17a) and $I_{out}$ (17b) are smaller than the critical current of the corresponding wires. For Nb wires with $\alpha j_c = 0.25$ A/μm², the minimum cross section of $L_q$ is $A_{min} = 8 \cdot 10^{-4}$ μm², which is an order of magnitude smaller than $A_{min}$ following from (14) and corresponds to wire dimensions ($A_{min})^{\frac{1}{2}} \sim 30$ nm. This is the ultimate limit to reduction of the linewidth of Nb wires.

Increasing $L_{ql}$ values significantly above 1 pH/μm is possible by using kinetic inductance (7). For instance, a 40-nm thick Mo₂N kinetic inductor in the SFQ5ee process [35] has $L_{Kl} \approx 8/w$ pH/μm (with $w$ in micrometers) [36], which is >>1 pH/μm at $w < 1$ μm. However, a short strip of kinetic inductor can be used only if $L_q$ is galvanically coupled to the next AQFP (so-called directly coupled AQFP [38]). If inductive (transformer) coupling to the next gate is used, the mutual inductance $M_q$ between the short strip of a kinetic inductor and the output inductor $L_{out}$, which is proportional to the length of $L_q$, is going to be small. Careful optimization is needed in this case, as will be discussed in Sec. III.

Consider inductive coupling with mutual inductance $M_q$ between the $L_q$ and the output inductor $L_{out}$ connected to the next AQFP or forming a part of the buffer which sums up the output currents of three AQFPs comprising the majority (MAJ3) gate [37]. Parametrically amplified input current $I_{in}$ creates current

$$I_q \approx \Phi_0/L_q = 2\pi I_{cJ}/\beta_q \approx 200 \text{ μA} \tag{17a}$$

in the $L_q$, which in turn creates current

$$I_{out} = M_q I_q/L_{out} \approx 2\pi M_q I_{cJ}/(\beta_q L_{out}) \tag{17b}$$

in the output inductor. The maximum value of $L_{out}$ can be determined from the condition $I_{out} \geq I_{in} \approx 10$ μA, which is the current required to drive the next AQFP. This gives

$$L_{out} \leq \frac{2\pi M_q I_{cJ}}{\beta_q I_{in}} = 19.63 M_q, \tag{18a}$$

$$l_{out} = \frac{19.63 M_q}{L_{outl}}. \tag{18b}$$

The latter determines the maximum distance over which the AQFP output data can be sent over without amplification, i.e., the maximum length of the inductor $L_{out}$; $L_{outl}$ is inductance per unit length of the output inductor. Using $M_{ql} \sim 0.3$ pH/μm, $l_q = 10$ μm, and $L_{outl} \sim 1$ pH/μm, we get $l_{out} \sim 60$ μm, about $6l_q$. Decreasing $L_{outl}$ by increasing the output inductor width may help in transmitting data to 10x larger distances than the height of the AQFP cell.

The $j_c$ for all kinetic inductors is much smaller than for Nb because of a much larger $\lambda$ in (13). For instance, if the $L_q$ is made of a 40-nm-thick Mo$_2$N kinetic inductor in the SFQ5ee process, its critical current would be reached at $w \leq 0.5$ μm [36]. Nevertheless, implementation of specially optimized kinetic inductors in AQFPs, and in superconductor electronics in general, could substantially decrease the size of logic cells and increase their number density [15].

### C. Cross Coupling of Transformers and Circuit Density

Another factor limiting the transformer and AQFP number density is cross-talk between different AQFPs. AQFPs are arranged along transmission lines (a power grid) feeding ac power into them [34], [37] as shown schematically in Fig. 1c. AQFPs arranged along one horizontal power line will also induce some ac excitation in the AQFPs coupled to the adjacent transmission lines of the gird. The minimum distance $S_{min}$ between the adjacent rows of AQFPs is determined by the acceptable level of cross-coupling between their transformers. We define this cross-coupling as the ratio of the mutual inductance $M_{cross}$ between the primary $L_1$ of the transformer in one row and the secondary $L_2$ of the transformer in the adjacent row to the mutual $M$ inductance within the same transformer. This ratio determines the ac excitation amplitude induced by a transmission line feeding one row of AQFPs into the adjacent rows of AQFPs below and above. In a given fabrication process, the minimum spacing between the adjacent rows of AQFPs is set by the max$[l_q, S_{min}]$, where $S_{min}$ is the spacing at which the maximum acceptable level of cross talk is reached. As an example, we will hereafter use 5% cross-talk as this maximum acceptable level.

Reducing the length $l_q$ below the $S_{min}$ would not increase $n_{QFP}$. This sets the minimum value of $L_{ql}$ required to maximize the density of AQFPs by using kinetic inductors

$$L_{ql} = \frac{\beta_{Lq}\Phi_0}{2\pi S_{min} I_{cJ}} \approx \frac{10.5}{S_{min}} \text{ (in pH/μm)}, \quad (19)$$

where $S_{min}$ is in micrometers.

We also need to consider cross-coupling between the adjacent AQFPs in the same row, i.e., coupling between parallel inductors $L_q$ and $L_{out}$ in two adjacent AQFPs. Current $I_{q1}$ in AQFP #1 induces flux $\Phi_{21} = M_{q21}I_{q1}$ in the adjacent AQFP #2 and current $I_{out21}$ in its output inductor $L_{2out}$. Reliable operation of the circuit may require this flux to be small in comparison with flux $\Phi_0$ created by ac excitation in AQFP #2 and the induced current to be small in comparison to self-current $I_{out22}$ (17) at the output of AQFP #2. As in the case of ac excitation transformers, we define cross coupling between

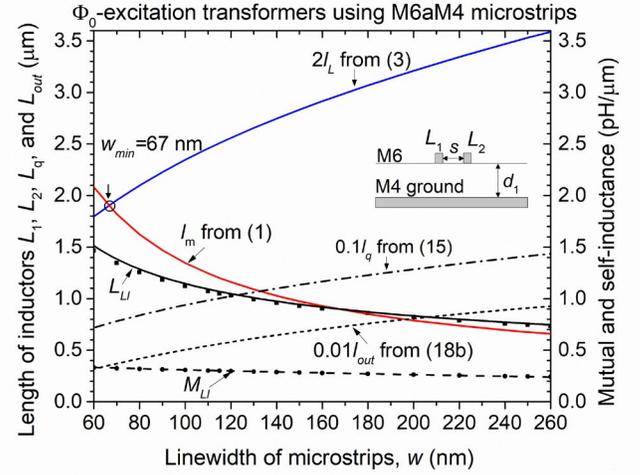

**Fig. 2.** Parameters of $\Phi_0$-excitation in-plane transformers using microstrips M6aM4 spaced at $s = 250$ nm in the SFQ5ee process; the transformers cross section is sketched in the Inset. Mutual inductance of the two microstrips per unit length, $M_l$ (right scale): black dash (21); (●) — numerical simulations using wxLC. Self-inductance per unit length, $L_{Ll}$ (right scale): black solid curve is the sum of magnetic (13) and kinetic (17) inductances; (■) — numerical simulations using wxLC. Thicknesses of Nb layers forming the microstrips: $t_{M6} = 200$ nm, M4 ground plane $t_{M4} = 200$ nm; dielectric thickness $d_1 = 655$ nm, $\lambda = 90$ nm. Solid red curve is the mutual running length $l_m$ (1) required to induce $\Phi_0$ in the AQFP secondary formed by two inductors $L_2$, assuming the maximum allowed current density of 0.25 A/μm² in the primary inductors $L_1$. The top solid blue curve is the length $2l_L$ (3) of the two inductors $L_2$ in the AQFP cell at $I_{cJ} = 50$ μA and $\beta_L = 0.2$. Below the $w_{min} \approx 70$ nm, corresponding to the circled intersection, $l_m > 2l_L$ and transformer with the required mutual inductance $M$ *cannot* be formed. Dash-dot curve is $0.1l_q$; $l_q$ is the length (15) of M6aM4 microstrip inductor $L_q$ with the same width as the $L_1$ and $L_2$; at $\beta_{Lq} = 1.6$, $l_q = 8l_L$. Short dash is $0.01l_{out}$; $l_{out}$ is the maximum length (18b) of the output inductor $L_{out}$, assuming it is also a M6aM4 microstrip with the same width $w$, coupled at $s = 250$ nm to the microstrip $L_q$ along the full length of the $L_q$.

the output transformers as

$$i_{21} = \frac{I_{out21}}{I_{out22}} = \frac{M_{q21}}{M_q}. \quad (20)$$

Its maximum tolerable value sets the minimum spacing $S_q$ between inductors $L_q$ and $L_{out}$ in the adjacent AQFPs in the same row along the ac power transmission line.

## III. Microstrip Transformers and AQFP Number Densities

For more accurate estimates of the number density of the ac transformers and AQFPs, in the following sections we will consider all types of possible transformers with the goal to minimize the area of the AQFP cell.

### A. AQFP Transformers Formed by Two Parallel Microstrips in One Plane

Mutual inductance of two microstrips in the same plane decreases slowly with distance between their centers $p_x = w + s$, and is given in [33] by

$$M_l = \frac{\mu\mu_0}{4\pi} \ln\left[1 + \frac{4\left(d_1 + \lambda + \frac{t}{2}\right)^2}{(w+s)^2}\right]. \quad (21)$$

For definiteness, we take parameters of the most advanced fabrication processes for superconductor electronics: the SFQ5ee process [35]. These parameters are given in Table I. Nb wires are on the process layer M6 with thickness $t = 200$ nm, 200-nm thick Nb ground plane is layer M4, and the dielectric thickness between them is $d_1 = 615$ nm. The currently allowed minimum linewidth and spacing between the microstrips in the same plane is 250 nm. Hopefully, with the progress of fabrication technology, narrower lines and gaps between metal lines with dielectric fill will become possible, e.g., due to the development of a damascene-type processing allowing for much smaller spacings $s \sim w$.

Fig. 2 shows the mutual inductance per unit length of two M6aM4 microstrips as a function of their linewidth at spacing $s = 250$ nm and the microstrip self-inductance per unit length (dash curve, scale on the right). It also shows the required mutual running length $l_m$ in (1) of the primary $L_1$, the length $2l_L$ in (3) of the secondary formed by two inductors $L_2$ (solid red curve). To show in the same scale, we also plotted $1/10^{th}$ of the $l_q$ given by (15) and $1/100^{th}$ of the $l_{out}$ given by (18b), assuming that the output transformer is also formed by two M6aM4 microstrips spaced at $s = 250$ nm, i.e., that the output transformer is of the same design as the ac excitation transformer. In addition to analytical expressions in [32], we used a very accurate numerical inductance extractor wxLC developed by M. Khapaev [38], assuming $\lambda = 90$ nm for all Nb layers. The numerical results are practically indistinguishable from the analytical ones.

Below the linewidth $w_{min} \approx 67$ nm, the required mutual running length of the excitation line $l_m$ needs to be longer than the total length $2l_L$ of the AQFP cell inductors $L_2$ in order to provide a $\Phi_0$ sum of the dc flux bias and ac excitation amplitude without exceeding the maximum allowed current $I_{max} = \alpha j_c w_{min} t_{M6} = 3.5$ mA in the primary. T*he parallel-line microstrip transformer with the required parameters cannot be formed using smaller linewidths.* The minimum cross section area of the primary $w_{min} t_{M6} = 0.0134$ μm² perfectly agrees with the $A_{min}$ estimated using (14) in II.A.

For hypothetical transformers with $s = w$, the excitation current limit is reached below $w_{min} \approx 50$ nm, corresponding to $I_{max} \approx 2.5$ mA, due to a stronger mutual coupling in the transformer.

In practical circuits, we want to reduce the ac excitation current $I_{ex}$ in the primary as much as possible in order to minimize ac power loss caused by dielectric losses in the transmission line, which grow as $I_{ex}^2$. The minimum $I_{ex}$ is obtained at the largest possible mutual running length at the given linewidth, i.e., at $l_m = 2l_L$. In this case, the full AQFP transformer consisting of two parallel wires in the same plane has area $A_{tr} = 2l_L(2w + s)$, where $l_L$ is given by (3). The minimum size of the rectangular AQFP cell for tiling is $(2l_L + s) \times (2w + 2s + l_q)$, where $(2l_L + s)$ is the minimum possible tiling pitch in the $x$-direction along the ac power transmission line, and $(2w + 2s + l_q)$ is the tiling pitch (effective height of the AQFP cell) in the $y$-direction.

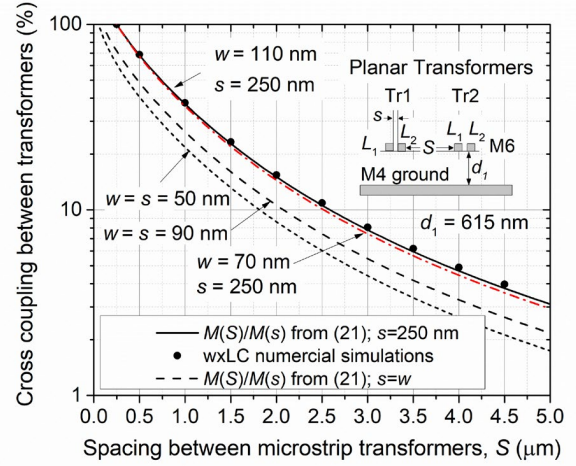

(a)

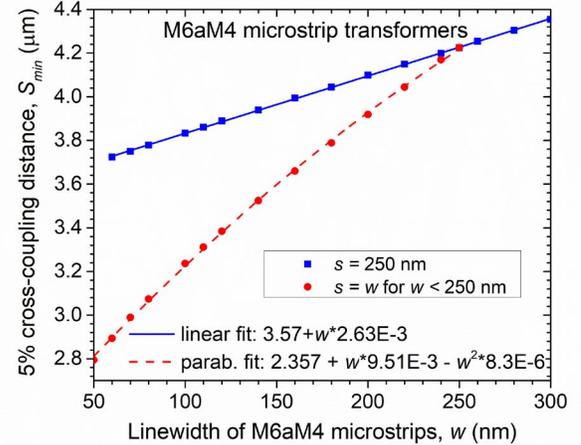

(b)

**Fig. 3.** (a) Cross coupling $M(S_y, w))/M(S, w)$ between in-plane (planar) microstrip-based transformers with various linewidths and in-transformer spacing $s$ as a function of spacing $S_y$ between the adjacent transformers. All the curves were calculated using (21) for the transformers using M6aM4 microstrips; ($\bullet$) – numerical simulations using wxLC software [38] for $w = 110$ nm.
(b) Dependence of the spacing $S_{min}$ at which mutual inductance between the transformers reduces to 5% of the in-transformer mutual inductance, referred to as a 5% cross-coupling distance, on the linewidth: blue squares – transformers with the primary to secondary spacing $s = 250$ nm; red dots – transformers in a hypothetical process with $s = w$ for $w \leq 250$ nm; solid blue line and red curve are, respectively, a linear and parabolic fits to the $S_{min}(w, s)$ in these two cases.

## B. Cross Coupling of Planar Microstrip-Based Transformers and the Limits on AQFP Number Density

To mitigate cross-coupling between two parallel transformers in the adjacent AQFPs, e.g., shown in Fig.1b, the spacing $S_y$ between the primary of transformer #1 and the secondary of transformer #2 should be much larger that the spacing $s$ between the wires in the transformer. The cross-coupling $\frac{M(S_y, w)}{M(s, w)}$ between two M6aM4 microstrip-based planar transformers is shown in Fig. 3a for a few linewidths, where $M(s,w)$ is given by (21) at $s = 250$ nm and $M(S_y, w)$ is given by (21) with $S_y$ replacing $s$. Numerical simulations using wxLC are shown by (●) and perfectly agree with (21).

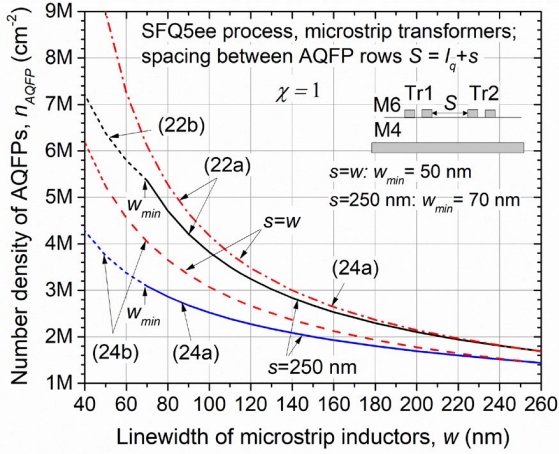

**Fig. 4.** Theoretical number density (assuming 100% area filling, $\chi = 1$) of AQFPs using M6aM4 microstrip inductors and transformers in the SFQ5ee process with parameters of Nb inductors $L_1$, $L_2$ and $L_q$ in Fig. 2 for the AQFP cell in Fig. 1. The bottom blue solid ($w > w_{min}$) and short dash ($w < w_{min}$) curves correspond to $s = 250$ nm and spacing $S_x$ between the AQFP in the $x$-direction set by a 5% cross coupling distance $S_q(w,s)$ because $2l_L + s < S_q(w,s)$ at all linewidths; spacing $S_y$ in the $y$-direction is set by the size of the AQFP cell, $l_q + s$. The black solid and short dash curves correspond to the $n_{AQFP}$ determined only by the physical dimensions of the AQFPs, respectively (22a) and (22b), ignoring cross-coupling. The upper red dash curve is (24) for a hypothetical process with $s = w$ and the physical dimensions of the AQFP, ignoring cross coupling between AQFPs in the same horizontal row. The red dash and short dash curves also correspond to $s = w$ but account for the cross-coupling in the same row (24b), using the dependence $S_q(w,s) = S_{min}(w,s)$ in Fig. 3b; see text. Cross-coupling between the adjacent rows in the $y$-direction can be neglected in all the cases because $l_q + s > S_{min}(w,s)$ at all the linewidths.

We see that providing a low crosstalk between the in-plane microstrip transformers requires quite large spacings, e.g., crosstalk below 10% requires $S_y > 2.5$ μm and below 5% requires $S_y \geq 3.75$ μm. The crosstalk increases with increasing $w$ and decreases with decreasing $s$.

The maximum acceptable crosstalk level depends on the circuit. For specificity, we hereafter take 5% as the maximum acceptable crosstalk level. This defines the minimum spacing between the transformers, $S_{min}(w,s)$, a function of the in-transformer linewidth and spacing, at which the 5% crosstalk is reached. This function is shown in Fig. 3b for the planar M6aM4 microstrip transformers with $s = 250$ nm (solid blue) in the SFQ5ee process and for the transformers with $w = s$ (red dash curve).

If we restrict the spacing between the horizontal rows of the AQFPs to the largest of the $l_q + s$ and $S_{min}$ in order to keep crosstalk between the ac excitation transformers at the acceptable level, the number density of the AQFPs becomes

$$n_{AQFP} = \frac{\chi}{(2l_L + s)(2w + 2s + l_q)}, \text{ if } l_q + s \geq S_{min}, w \geq w_{min}, \quad (22a)$$

where $\chi \leq 1$ is the area filling factor.

Linewidths $w < w_{min}$ cannot be used in the AQFP ac excitation transformers because of the critical current limitation, *but can be used for the inductor $L_q$*. In this case, the cell height continues to decrease because reduction of the inductor $L_q$ width reduces its length $l_q$, whereas the cell effective width $(2l_L + s)$ remains constant and equal $2l_L(\text{at } w_{min}) + s$. Hence, if $w < w_{min}$ and $l_q + s \geq S_{min}$, the number density of AQFPs is given by

$$n_{AQFP} = \frac{\chi}{\left(2l_L|_{w=w_{min}} + s\right)(2w_{min} + 2s + l_q)}. \quad (22b)$$

In the opposite case $S_{min} > l_q + s$, the number density is given by

$$n_{AQFP} = \frac{\chi}{(2l_L + s)(2w + s + S_{min})}, \text{ if } S_{min} \geq l_q + s, w \geq w_{min} \quad (23a)$$

$$n_{AQFP} = \frac{\chi}{\left(2l_L|_{w=w_{min}} + s\right)(2w_{min} + s + S_{min})}, \text{ if } S_{min} \geq l_q + s \text{ and } w < w_{min}. \quad (23b)$$

Using Nb microstrips M6aM4 as AQFP inductors in the SFQ5ee process, gives $l_q \geq S_{min}$ for all linewidths down to about 20 nm. Hence, (22a) applies if we ignore cross coupling between adjacent AQFPs in the same row; see below. Dependences (22a) and (22b) for AQFPs with the M6aM4 inductors and transformers with $s = 250$ nm are shown in Fig. 4 by the solid black ($w \geq w_{min}$) and black dash ($w < w_{min}$) curves.

So far, we have ignored crosstalk between the data output transformers $L_q$, $L_{out}$ of adjacent AQFPs in the same horizontal row. However, placing adjacent AQFPs in the same row at the minimum spacing $s = 250$ nm gives the horizontal spacing $S_x$ between inductors $L_q$ in the range $2.75 \leq 2l_L + s \leq 3.85$ μm. At these distances, the crosstalk between the parallel output inductors can be very substantial as follows from (21) and Fig. 3. So, the AQFPs may need to be placed further apart in the $x$-direction (along the ac transmission line) and spaced at some distance $S_q$ to reduce the output crosstalk to the acceptable level. Then, (22a) and (22b) are replaced by, respectively, (24a) and (24b)

$$n_{AQFP} = \frac{\chi}{(S_q + w)(2w + 2s + l_q)}, \text{ if } l_q + s \geq S_{min}, S_q > 2l_L + s, \text{ and } w \geq w_{min} \quad (24a)$$

$$n_{AQFP} = \frac{\chi}{(S_q + w)(2w_{min} + 2s + l_q)}, \text{ if } w < w_{min}, \text{ and } l_q + s \geq S_{min}, S_q > 2l_L + s. \quad (24b)$$

If the minimum spacing of AQFPs in both the $x$- and $y$-directions is set by the acceptable cross-coupling, then

$$n_{AQFP} = \frac{\chi}{(S_q + w)(2w + s + S_{min})}, \text{ if } l_q + s < S_{min}, \ S_q > 2l_L + s, \text{ and } w \geq w_{min}, \quad (25a)$$

$$n_{AQFP} = \frac{\chi}{(S_q + w)(2w_{min} + s + S_{min})}, \text{ if } l_q + s < S_{min}, \ S_q > 2l_L + s, \text{ and } w < w_{min}. \quad (25b)$$

If the acceptable cross-talk level between the adjacent inductors $L_q$ is the same as between the excitation transformers, e.g., less than 5%, then $S_q = S_{min}$ because the output inductors are of the same type as in the ac excitation transformers. Then, the minimum pitch of AQFP placement in the same row is $S_{min} + w$, and the number density of AQFPs using the planar M6aM4 transformers is given by (24a) and (24b). The corresponding number density of AQFPs is shown in Fig. 4 by the lowest blue solid (24a) and short-dash (24b) curves calculated using the dependences in Fig. 3b and $S_q(w,s) = S_{min}(w,s)$.

For the currently available SFQ5ee process with $s = 250$ nm and already demonstrated linewidth $w \approx 110$ nm [10], [11], the theoretical AQFP number density is about $3.5 \cdot 10^6$ cm$^{-2}$, set by the physical dimensions of the AQFP and ignoring in-row cross-talk. It reduces to about $2.5 \cdot 10^6$ cm$^{-2}$ if crosstalk needs to be kept below the 5% level. These densities correspond to, respectively, 1.17M and 0.83M MAJ3 logic gates per cm$^2$. We emphasize that the considered AQFPs are about 30 times smaller in area than 30 μm x 40 μm AQFP buffer cells used in [37].

Reducing spacing between superconducting lines below 250 nm in a hypothetical future development of the SFQ5ee process would bring some increase in the device density, as follows from (21)–(24) and Fig. 3b. For instance, using $s = w$ would increase mutual coupling in the planar transformer, resulting in $w_{min} \approx 50$ nm. Cross-coupling between the microstrip transformers would also noticeably decrease, resulting in $S_q(5\%) = S_{min}(5\%) \approx 2.75$ μm at $w = 50$ nm, as shown in Fig. 3b. This number can be used to estimate the absolutely maximum possible value of $n_{AQFP}$ in this technology.

Even at $w = 40$ nm, $l_q = 5.66$ μm is still twice as large as the $S_{min}$, and the $n_{AQFP}$ is still given by (22a) and (22b), as shown in Fig. 4 by the upper dash-dot red curve, reaching about $9 \cdot 10^6$ cm$^{-2}$ at $w = 50$ nm. If the placement pitch $S_x$ of AQFPs in the same row is set by the 5% crosstalk requirement between the output transformers $S_q = S_{min}(w,s)$, the $n_{AQFP}$ is given by (24a) and (24b), and shown in Fig. 4 by the red dash ($w \geq w_{min}$) and short dash ($w < w_{min}$) curves, respectively, reaching about 6.2M cm$^{-2}$ at $w = 40$ nm.

For a more realistic 90-nm process ($w = s = 90$ nm), $n_{AQFP} = 4.65 \cdot 10^6$ cm$^{-2}$ and $3.80 \cdot 10^6$ cm$^{-2}$ without and with accounting for the in-row cross-talk, respectively. This is less than 33% increase over the density of AQFPs at $w = 90$ nm and $s = 250$ nm, which would hardly justify a very complex process development required for achieving the much smaller spacing $s = 90$ nm.

### C. AQFPs with M6aM4 Microstrips and Kinetic Inductor $L_q$

The main density limiter of AQFPs in the SFQ5ee process is a very large length of the inductor $L_q$. Let us assume that in a hypothetical, next generation process, we can introduce an additional layer of kinetic inductors close to the layer of JJs and use it to make $l_q + s \leq S_{min}(w,s)$, the length set only by the crosstalk requirements. This would bring the AQFPs in the regime (23) or (25). Using (19), the optimum value of the linear inductance of the kinetic inductance material replacing Nb in the M6aM4 microstrips for $L_q$ would be $L_{ql} \approx 2.8$ pH/μm, a factor of 4 higher than the linear inductance of 250-nm-wide Nb microstrips M6aM4. For the reasonable width of the $L_q$, $w_q = 250$ nm, the required sheet inductance is 0.7 pH/sq, a factor of 10 lower than the sheet inductance of the 40-nm Mo$_2$N films [36] currently used in the SFQ5ee process as rf choke kinetic inductors for biasing ERSFQ circuits [40].

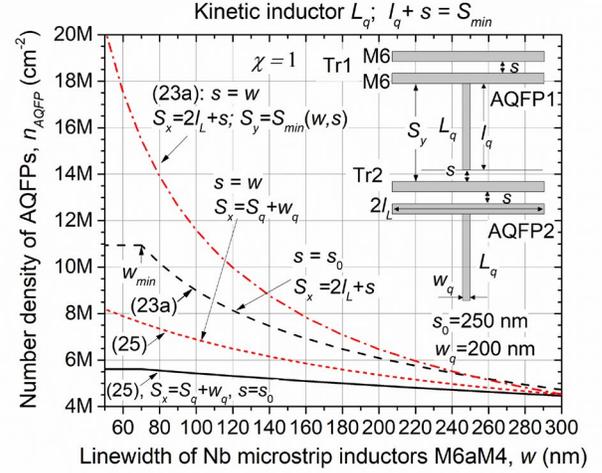

**Fig. 5.** Theoretical number density (assuming 100% area filling, $\chi = 1$) of AQFPs $n_{AQFP}$ in a hypothetical process using Nb microstrip inductors M6aM4 in the ac excitation transformers and a kinetic inductor $L_q$ with width $w_q = 200$ nm and length $l_q + s = S_{min}(w,s)$, where $S_{min}(w,s)$ is a 5% cross-coupling spacing between the AQFP primary in one row and the AQFP secondary in the adjacent row shown in Fig. 3b. The bottom solid black curve (25) corresponds to $s = 250$ nm and the AQFP placement pitch in the $x$-direction equal to $S_x = S_q(w_q) + w_q$ determined by the 5% cross-coupling level between inductors $L_q$ and $L_{out}$ in the adjacent AQFPs in the same horizontal row. The dash black curve (23) corresponds to the $n_{AQFP}$ at $S_x$ equal the physical width of the ac transformer $2l_L + s$ and $S_y = S_{min}(w,s)$. Red dash-dot curve (23a) corresponds to a hypothetical process with spacing $s = w$ in the excitation transformer and the AQFP cell size $2l_L + s$ determined by the transformer length; the number density saturates below $w_{min} = 50$ nm at $n_{AQFP} = 20.2$M per cm$^2$, a level given by (23b). Red short-dash curve also corresponds to the case $s = w$, but with spacing between the inductors $L_q$ set by a 5% cross-coupling level $S_q(w_q) = 3.92$ μm according to the data in Fig.

If in the hypothetical fabrication process we preserve somehow Nb microstrips M6aM4 in the ac excitation transformer, the $n_{AQFP}$ could be increased up to the limit set only by the acceptable cross-talk between the AQFPs. For this, the pitch of the AQFP placement in $y$-direction should be $2w + s + S_{min}$, and the pitch in the $x$-direction should be $S_x = S_q + w_q$, where $w_q$ is the width of the kinetic inductor $L_q$, because in the entire range of linewidths in Fig. 4, the physical width of the AQFP transformers $2l_L + s$ is less than $S_q(5\%)$. Due to the critical current limitation, the minimum expected value of $w_q$ is about 200 nm. In the described case, the expected number density of AQFPs using a kinetic inductor $L_q$ depends very weekly on the $w$ and $w_q$ as can be seen from the lowest black curve in Fig. 5. Below $w_{min} = 50$ nm, $n_{AQFP}$ in this regime does not increase because the height of the AQFP cell is set by the $S_{min}(w_{min}, s)$ and no longer depends on the linewidth. The saturation value $(n_{AQFP})_{max} = 5.6 \cdot 10^6$ cm⁻² is larger than the number density which could be achieved using the ultranarrow Nb wires; see the bottom blue curves in Fig. 4. More importantly, these densities can be achieved at modest and already demonstrated linewidths and spacings, and only require implementing a kinetic inductor for $L_q$.

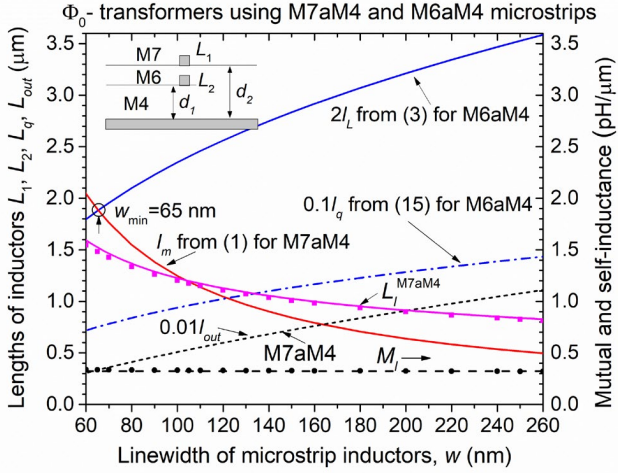

**Fig. 6.** Parameters of the ac excitation transformers using aligned equal-width microstrips M7aM4 and M6aM4 for the transformers primary and secondary, respectively; see the Inset: top blue curve – total length $2l_L$ (3) of two AQFP inductors $L_2$ (microstrips M6aM4) forming the transformer secondary, at $\beta_L = 0.2$, $I_{cJ} = 50$ μA; red solid curve – mutual running length $l_m$ in (1) of the transformer primary (microstrip M7aM4) and the secondary required to provide a $\Phi_0$ flux excitation (dc + ac) in the secondary at the excitation current $I_{max}$ (9) in the primary, assuming the maximum current density in Nb of 0.25 A/μm². Below $w_{min} \approx 65$ nm, $l_m > 2l_L$ and the required flux excitation cannot be provided without exceeding the critical current of Nb wire M7 in the transformer primary. Mutual inductance between the M7aM4 and M6aM4 microstrips per unit length $M_l$ calculated using (26) at $d_1 = 615$ nm, $d_2 = 1015$ nm, $\lambda = 90$ nm, and $t_{M4} = t_{M6} = t_{M7} = 200$ nm is shown by the bottom black dash line; ($\bullet$) – numerical simulations using wxLC and the same parameters. Self-inductance of M7aM4 microstrips per unit length, $L_l^{M7aM4}$: solid magenta curve – sum of (13) and (7); (■) –numerical simulations using wxLC. Dash-dot blue curve is $0.1l_q = 0.8l_L$ (at $\beta_q = 1.6$) calculated using (15) and assuming that inductor $L_q$ is a M6aM4 microstrip of the same width as the $L_1$ and $L_2$. Black short-dash curve is the 1/100th of the length of inductor $L_{out}$ calculated using (18b) and $M_q = M_l$, assuming it is a M7aM4 microstrip aligned over the $L_q$ and having the of the same linewidth $w$.

If a more aggressive process with $s = w$ is used, even higher values of $n_{AQFP}$ could be achieved as shown in Fig. 5 by the upper dash-dot curve, corresponding to (23a), because of a smaller cross-talk and smaller $S_{min}$ between the ac transformers; see Fig. 3b. At $w < w_{min}$, the $n_{AQFP}$ saturates at $(n_{AQFP})_{max} = 2 \cdot 10^7$ cm⁻², a level given by (23b) at $w = w_{min} = 50$ nm. This packing density could provide up to 6.7M MAJ3 logic gates per cm² and *would be an 8x improvement over the standard SFQ5ee process.* However, it may still be not sufficient for general purpose computing applications.

It appears that no significant increase in the circuit density is possible with microstrip inductors in planar transformers, mainly because of their strong cross-coupling. Accounting for other omitted components, e.g., JJs and interlayer vias, may only reduce the maximum densities estimated above. For the sake of completeness, we will consider transformers based on mutual coupling of vertically spaced microstrips in the next section, and then turn to stripline-based transformers in Sec. IV.

### D. AQFP Transformers Formed by Aligned Microstrips on the Vertically Spaced Planes

The area of AQFP ac excitation transformers can be decreased, at least by a factor of three, if microstrips forming the transformer have the same width and locate above each other, over the same ground plane. In this case $A_{tr} = 2l_L w$.

If $L_2$ of the AQFP is the M6aM4 microstrips considered in Secs. III.*A* and III.*B*, the only convenient transformer primary is a microstrip M7aM4 with the signal trace on niobium layer M7. Mutual inductance per unit length of two microstrips with widths $w_1$, $w_2$, and thicknesses $t_1$, $t_2$ located on different planes is given in [33] by

$$M_l = \frac{\mu\mu_0}{4\pi}\ln\left[1 + \frac{4\left(d_1 + \lambda + \frac{t_1}{2}\right)\cdot\left(d_2 + \lambda + \frac{t_2}{2}\right)}{p_x^2 + (d_2 - d_1)^2}\right], \quad (26)$$

where $d_2$ and $d_1$ are the dielectric thicknesses between the respective signal traces and the ground plane, and $p_x = s + \frac{w_1 + w_2}{2}$ is the horizontal distance between the geometrical centers of the microstrips' cross sections. In the SFQ5ee process, $t_1 = t_2 = 200$ nm for both layers M6 and M7, $d_1 = 615$ nm, and $d_2 = 1015$ nm, corresponding to the interlayer dielectric thickness of 200 nm.

Mutual inductance (26) of the aligned microstrips M7aM4 and M6aM4, $p_x = 0$, is shown in Fig. 6 by the lowest black dash line along with the numerically simulated dependence shown by solid dots ($\bullet$). Self-inductance per unit length of M7aM4 microstrips is given by the sum of (13) and (7). Similarly to the planar transformers in Fig. 2, the required mutual running length of wires $l_m$ in the "vertical" transformer needs to become longer than two AQFP inductors $L_2$ (the transformer secondary) at $w_{min} \lesssim 65$ nm because the sum of the ac and dc currents in the primary required to induce flux $\Phi_0$ in the AQFP reaches the maximum allowed value $I_{max} = \alpha j_c w_{min} t = 3.25$ mA. The required transformer cannot be formed using narrower wires. The minimum cross section $w_{min} t_{M7} = 0.013$ μm² agrees perfectly with $A_{min}$ estimated in II.A from the solution of (14).

Crosstalk between two parallel vertical M7-M6aM4 transformers is shown in Fig. 7 as a function of spacing $S$ between them, for two linewidths $w = 250$ nm and 65 nm. Crosstalk is defined as the ratio of the mutual inductance between the M7 wire in transformer #1 and the M6 wire in transformer #2 to the mutual inductance in the transformer. The crosstalk very weakly depends on the $w$; it reduces below the 5% level at $S \geq S_{min} = 4.62$ μm and 4.44 μm, respectively, at $w = 65$ nm and 250 nm. Hence, instead of using a function $S_{min}(w)$ as in III.B, we can simply use a single averaged value $S_{min} = 4.53$ μm in the entire range of linewidths of interest. Overall, the crosstalk between the vertical transformers is larger than between the planar M6-M6aM4 transformers in Fig. 3 because $d_2$ is significantly larger than $d_1$.

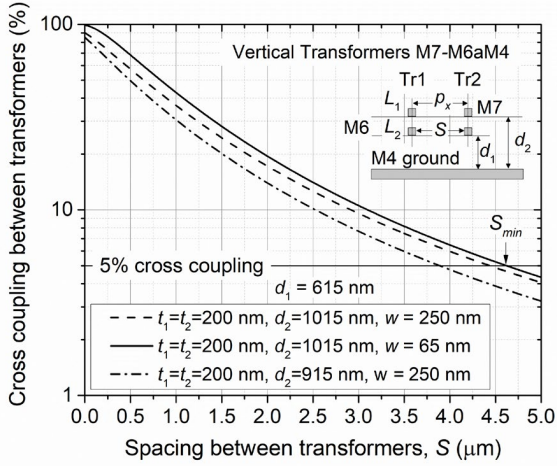

**Fig. 7.** Crosstalk between vertical transformers formed by the aligned equal-width M7aM4 and M6aM4 microstrips as a function of spacing between the transformers. Crosstalk is defined as mutual inductance between the primary of transformer #1 (Tr1) and the secondary of transformer #2 (Tr2) normalized to the mutual inductance inside the transformer. All curves were calculated using (26): solid curve corresponds to the parameters of the existing SFQ5ee process and linewidth $w$ =250 nm; dash curve is for $w = w_{min}$ =65 nm; dash-dot curve corresponds to a hypothetical process with a reduced dielectric thickness between M6 and M7. $S_{min}$ is defined as the spacing at which the tolerable level of cross coupling, set as 5% in this case, is reached. Dependence of $S_{min}$ on $w$ is very weak and an average value $S_{min}$ =4.53 µm will be used to characterize the entire range of linewidths of interest $w_{min} \leq w \leq 300$ nm.

If $L_q$ is an M6aM4 microstrip with the same width as the $L_2$, its length $l_q$ is the same as was described in III.B, Fig. 2, and $l_q > S_{min}$. Hence, the AQFP number density only slightly differs from (22) and, neglecting cross-coupling in the same row of AQFPs, is

$$n_{AQFP} = \frac{\chi}{(2l_L + s)(w + l_q + s)}, \text{ if } l_q + s \geq S_{min}, 2l_L + s \geq S_q. \quad (27a)$$

This dependence is shown in Fig. 8 by the solid blue curve in the middle. At $w = 65$ nm and $s = 250$ nm, (27a) gives $n_{AQFP} \approx 6 \times 10^6$ cm$^{-2}$, corresponding to about 2M MAJ3 gates per cm$^2$.

Using $w < w_{min}$ for the inductor $L_q$ reduces its length $l_q$. The AQFP number density in this case is similar to (22b) and given by

$$n_{AQFP} = \frac{\chi}{(2l_L|_{w=w_{min}} + s)(w_{min} + l_q + s)}, \text{ if } w < w_{min}, l_q + s \geq S_{min}, 2l_L|_{w=w_{min}} + s \geq S_q. \quad (27b)$$

This dependence is shown in Fig. 8 by the blue dash curve.

If cross-coupling between the AQFP outputs needs to be kept below a certain level, the adjacent inductors $L_q$ in the same row need to be spaced at a safe distance $S_q$, and the AQFP number density reduces to

$$n_{AQFP} = \frac{\chi}{(s_q + w)(w + l_q + s)}, \text{ if } w \geq w_{min}, l_q + s \geq S_{min}, S_q + w \geq 2l_L + s. \quad (28a)$$

$$n_{AQFP} = \frac{\chi}{(s_q + w)(w_{min} + l_q + s)}, \text{ if } w < w_{min}, l_q + s \geq S_{min}, S_q + w > 2l_L + s. \quad (28b)$$

These dependences are shown by the lowest black solid and dash curves in Fig. 8 for a 5% cross-coupling requirement $S_q = S_{min}$ =4.53 µm, following from Fig. 7.

Since for all considered linewidths, $l_q + s > S_{min}$, the size of the AQFP cell can be reduced by using a kinetic inductor for $L_q$ to make $l_q + s \leq S_{min}$. In this hypothetical process, the number density of AQFPs is similar to (23) and given by

$$n_{AQFP} = \frac{\chi}{(2l_L + s)(w + S_{min})}, \text{ if } S_{min} \geq l_q + s, w \geq w_{min}, 2l_L + s \geq S_q + w \quad (29a)$$

$$n_{AQFP} = \frac{\chi}{(2l_L|_{w=w_{min}} + s)(w_{min} + S_{min})}, \text{ if } w < w_{min}, S_{min} > l_q + s, 2l_L|_{w=w_{min}} + s \geq S_q + w. \quad (29b)$$

These dependences are shown in Fig. 8 by the red solid curve saturating at $(n_{AQFP})_{max}$ =1.02·10$^7$ cm$^{-2}$, according to (29b). And, finally, when the crosstalk is the factor in choosing the $x$- and $y$- placement pitches, the number density is given by

$$n_{AQFP} = \frac{\chi}{(s_q + w)(w + S_{min})}, \text{ if } S_{min} \geq l_q + s, S_q + w_q > 2l_L + s, w \geq w_{min} \quad (30a)$$

and reaches the constant density at $w < w_{min}$

$$n_{AQFP} = \frac{\chi}{(s_q + w_q)(w_{min} + S_{min})}, \text{ if } S_{min} \geq l_q + s, S_q + w_q >$$

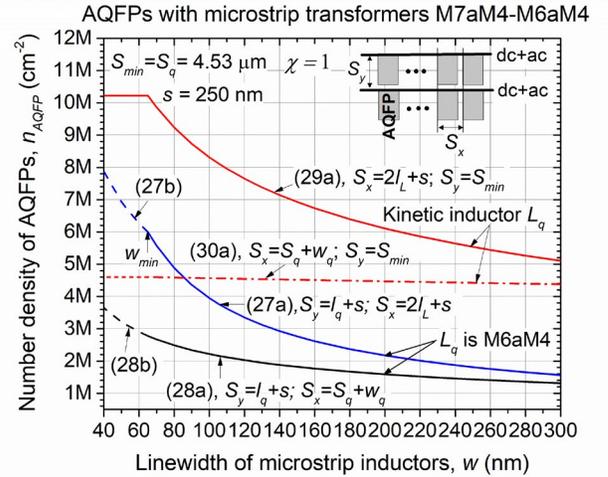

**Fig. 8.** Theoretical number density of AQFPs using vertical transformers formed by aligned, equal-width microstrips M6aM4 and M7aM4 as a function of their linewidth. The bottom black solid and dash curves correspond to Nb microstrips in the SFQ5ee process and the AQFP cell spacing $S_y = l_q + s$ (see the Inset) set by the physical height of the cell and the cell effective width $S_x = S_q + w$ set by the 5% crosstalk spacing $S_q$ equal approximately 4.53 µm in the entire range of the linewidths. Blue solid and dash curves also correspond to Nb microstrips, but the cell placement pitch set by the physical width $S_x = 2l_L + w$, ignoring the crosstalk. The upper red solid and dash-dot curves correspond to a hypothetical process, in which AQFP inductor $L_q$ on the M6 level is made of a kinetic inductance material with width $w_q$ =200 nm to reduce its length below the critical cross-coupling distance $S_{min}$ while preserving Nb microstrips in the ac excitation transformer.

$$2l_L + s, w < w_{min}. \quad (30b)$$

These dependences are shown in Fig. 8 by the nearly horizontal red dash-dot and short-dash lines. The maximum number density in (30b) is $(n_{AQFP})_{max} = 4.4 \cdot 10^6$ cm$^{-2}$ at the kinetic inductor $L_q$ width $w_q = 0.2$ μm. This is a bit lower number than for the planar M6-M6aM4 transformers because of a larger $S_{min}-$ the spacing at which the cross coupling reduces to below 5%.

The $n_{AQFP}$ can be increased in a hypothetical fabrication process by using smaller thicknesses $t_1$, $t_2$, and $d_2$ to increase the mutual inductance in the transformer and decrease cross coupling. For instance, at $t_1 = t_2 = 200$ nm and $d_2 = 915$ nm, corresponding to the interlayer dielectric thickness between the M7 and M6 wires of 100 nm and $\Delta d = d_2 - d_1 = 300$ nm, the $S_{min}$ would decrease to about 3.8 μm, as shown in Fig. 7.

To conclude, the achievable number densities of AQFPs using vertical microstrip-based transformers are comparable to those obtainable with the planar microstrip transformers in III.B because both types of the transformers have close values of the mutual inductance and similarly strong cross coupling – the main drawback of using microstrips.

## IV. STRIPLINE TRANSFORMERS AND AQFP NUMBER DENSITIES

### A. Stripline Transformers in the SFQ5ee Process

It is well known that cross talk can be substantially reduced using stripline inductors with two ground planes instead of microstrips with one ground plane. The mutual inductance per unit length between two superconducting striplines is given in [33] by

$$M_l = \frac{\mu_0}{4\pi} \ln \frac{\cosh\frac{\pi p_x}{H+2\lambda} - \cos\frac{\pi(h_1+h_2+2\lambda)}{H+2\lambda}}{\cosh\frac{\pi p_x}{H+2\lambda} - \cos\frac{\pi(h_2-h_1)}{H+2\lambda}}, \quad (31)$$

where $H$ is the distance between the ground planes and $h_i = d_i + \frac{t_i}{2}$ is the distance between the bottom ground plane and the geometrical center of the cross section of the $i$-th signal wire. At large distances, $M_l$ exponentially decreases with the in-plane (horizontal) distance $p_x$ between the geometrical centers of the stripline cross sections, with a decay length $p_0 = \frac{H+2\lambda}{\pi}$.

In the existing SFQ5ee process, two parallel M6aM4bM7 (standing for M6 above M4 below M7) striplines can be used for a planar stripline-based transformer near the JJs. In this case, $H = 1015$ nm and $p_0 = 380$ nm. Other possibilities would involve layers below the JJ and require multiple vias, leading to larger cell areas.

Because of a lower self-inductance and three times lower mutual inductance of the M6aM4bM7 striplines compared to the M6aM4 microstrips, lengths of the inductors $L_2$ and $L_q$, and the minimum width $w_{min} = 105$ nm, shown in Fig. 9, are noticeably larger in this case than in Fig. 2. This leads to larger AQFP cell sizes and smaller $n_{AQFP}$ shown by a solid black curve in Fig. 10. However, *the cross-coupling between the striplines in the adjacent transformers is dramatically lower than between the microstrips.* Assuming a purely exponential

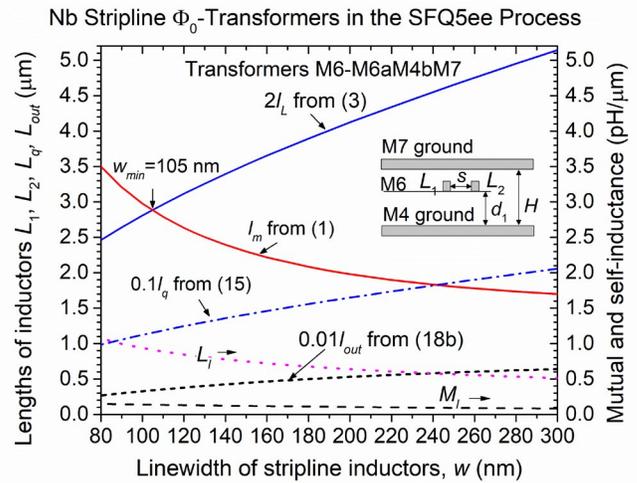

**Fig. 9.** Parameters of $\Phi_0$-exitation transformers using M6aM4bM7 striplines spaced at $s = 250$ nm (see sketch in the Inset) in the SFQ5ee process with parameters: $d_1 = 615$ nm, $H = 1015$ nm, $\lambda = 90$ nm, $t_{M4} = t_{M6} = t_{M7} = 200$ nm. Solid red curve – mutual coupling length $l_m$ in (1) of the primary $2L_1$ and the secondary $2L_2$ as a function of their linewidth; solid blue curve - length $2l_L$ of two inductors $L_2$ in (3) at $\beta_L = 0.2$, $l_{cJ} = 50$ μA. The magenta dot curve is self-inductance of the M6aM4bM7 striplines per unit length, $L_l$, the scale is on the right axis; the bottom black dash curve is mutual inductance $M_l$ per unit length between the striplines at $s = 250$ nm. The blue dash-dot curve is $l_q/10$ from (15), which is equal $0.4l_L$, assuming that inductor $L_q$ is a M6aM4bM7 stripline of the same linewidth as the $L_1$ and the $L_2$. The short dash black curve is $l_q/100$ calculated using (18b) and $M_q = M_l$, assuming that inductor $L_{out}$ is also a M6aM4bM7 stripline spaced from the $L_q$ at 250 nm. Below $w_{q} \approx 105$ nm, the primary current inducing the $\Phi_0$ excitation at the coupling length $l_m = 2l_L$ exceeds 5.25 mA, the $I_{max}$ of Nb wires. The assumed maximum excitation current density in the primary stripline is 0.25 A/μm$^2$.

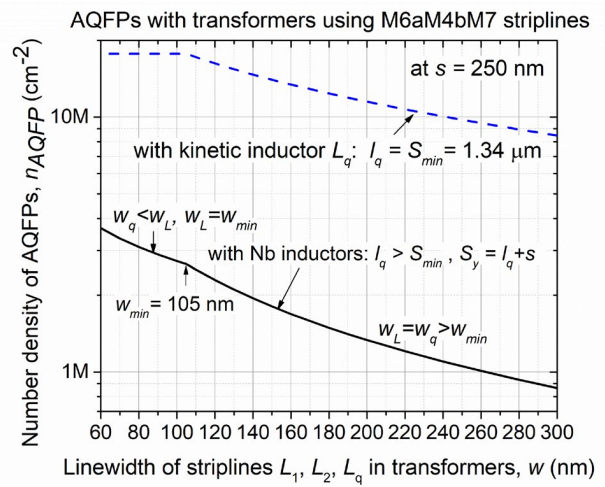

**Fig. 10.** Number density of AQFPs using striplines M6aM4bM7 spaced at $s = 250$ nm in all transformers: lower solid black curve – Nb striplines in the SFQ5ee process with M4 and M7 ground planes at $H = 1015$ nm; blue dash curve – a hypothetical process with Nb striplines in the ac excitation transformer $L_1$, $L_2$, and a kinetic inductor $L_q$ with length equal to the 5% cross talk length $S_{min} = 1.34$ μm at $w = w_{min} = 105$ nm.

decay, a 5% cross coupling is expected at $S_{min} \approx p_0 |\ln 0.05| = 1.14$ μm. The cross-coupling calculated using the full expression (31) at the in-transformer spacing $s = 250$ nm

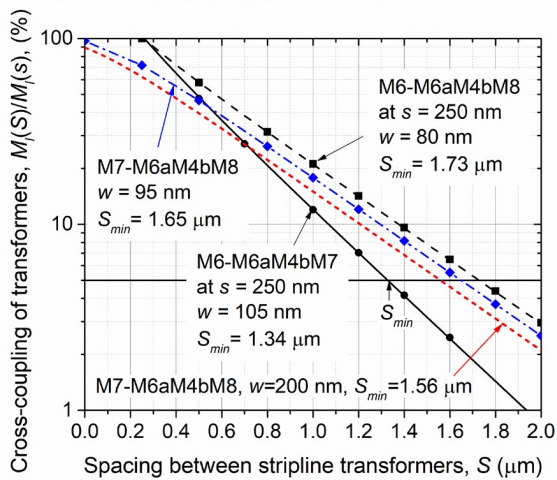

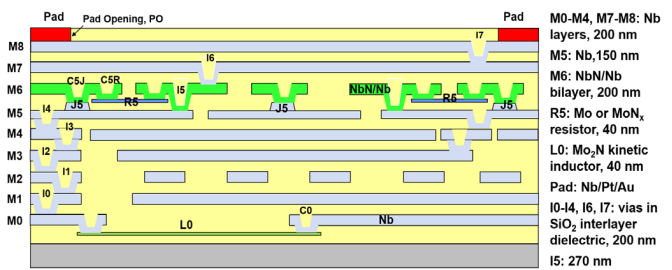

**Fig. 12.** Cross section of an advanced process node SFQ7ee with 10 superconducting layers. In comparison to the standard node SFQ5ee, it has an additional Nb layer M8. The layer M6 can be deposited either as a 200-nm Nb layer or as a NbN/Nb bilayer, e.g., with equal 100 /100 nm thicknesses. In the latter case, 100-nm thick NbN kinetic inductors can be formed by selectively etching the top Nb layer of the bilayer, while the full bilayer can be used to form regular, lower-value, geometrical inductors as well as ac power and data transmission lines.

**Fig. 11.** Cross coupling between stripline transformers as a function of spacing $S$ between them: (●) - transformers formed by two striplines M6aM4bM7 spaced at $s =$250 in the SFQ5ee process using M4 and M7 ground planes at $H =$1015 nm, $w = w_{min} =$105 nm; (■) - transformers formed by two striplines M6aM4bM8 spaced at $s =$250 nm in the SFQ7ee process using M4 and M8 ground planes at $H =$1415 nm, $w = w_{min} =$80 nm, (♦) - transformers formed by a stripline M7aM4bM8 (dielectric thickness between M4 and M7 $d_2 =$1015 nm, the SFQ7ee process) aligned over a stripline M6aM4bM8; width of both striplines $w = w_{min} =$95 nm; red short dash is for the same striplines but with $w =$200 nm. All the shown curves were calculated using (31), points are numerical simulation using wxLC. In all the cases, we used $\lambda =$ 90 nm, and thicknesses of the Nb layers $t_{M4} = t_{M6} = t_{M7} = t_{M8} =$200 nm.

and numerical simulations give $S_{min} =$1.34 μm; see Fig. 11.

The number density of AQFPs using M6aM4bM7 striplines in Fig. 10 is given by (22a) and (22b) because $l_q \gg S_{min}$ and $2l_L + s > S_{min}$ in the entire range of $w$. Using $w < w_{min}$ is possible for the inductor $L_q$, while keeping the excitation transformer linewidth at $w_{min}$; this reduces the length $l_q$ and the AQFP cell area. At $w = 60$ nm, $n_{AQFP} \approx 3.7 \times 10^6$ cm$^{-2}$ and limited only by the cell size and not by the cross-talk. This is slightly lower a number than one can get using the M6aM4 microstrip inductors, for which the AQFP number density is limited by the cross-talk; see the lowest curve (labeled 24b) in Fig. 4.

If using a kinetic inductor we could reduce length of the inductor $L_q$ to $l_q = S_{min} =$1.34 μm, the maximum number density would increase dramatically to $n_{AQFP} \approx 1.8 \times 10^7$ cm$^{-2}$, as shown in Fig. 10, whereas the in-row and between-rows cross-talk would remain below 5%. In the next section we will consider the fabrication process required to achieve this.

## V. Advanced Fabrication Process SFQ7ee for Superconductor Logic/Memory Using AC Excitation

### A. Nb Stripline Transformers in the SFQ7ee Optimized Process

In order to utilize advantages of a very low cross-talk between stripline transformers (small $S_{min}$) and substantially increase the number density of logic cells using flux transformers and ac excitation, we need to engineer an optimized fabrication process. From (28), the mutual inductance is maximized by decreasing the vertical distance between the signal wires $h_2 - h_1$ and placing them such that $h_2 + h_1 = H$, i.e., symmetrically in the middle between the ground planes at around $\frac{H}{2}$. These conditions could be nearly satisfied if we add a niobium layer M8, above the layer M7 in the existing SFQ5ee process. This would allow us to use M6aM4bM8 striplines in planar transformers and M7aM4bM8 striplines in vertical transformers. If we use 200 nm interlayer dielectric I7 (between M7 and M8), the distance between the ground planes M4 and M8 would become $H =$1415 nm, giving the decay length $p_0 = 508$ nm. The corresponding fabrication process, titled SFQ7ee, is currently under development at MIT LL. Its cross section is shown in Fig. 12.

Self- and mutual inductances of Nb striplines M6aM4bM8 and M7aM4bM8 in the SFQ7ee process are shown in Fig. 13a. Results of the analysis of transformers in the SFQ7ee process is shown in Fig. 13b. For the planar M6-M6aM4bM8 excitation transformers in AQFPs we get: $w_{min} \approx$80 nm at spacing between the M6 strips $s =$250 nm and $w_{min} \approx$55 nm in the hypothetical case of $s = w$. In the former case, at $w =$80 nm, $l_L =$1.52 μm, $l_q \approx$12 μm, and a 5% cross-talk is reached at $S_{min} =$1.73 μm; see Fig. 11. We note that $S_{min}$ slightly decreases with increasing the linewidth. Hence, we used the largest $S_{min}$ corresponding to the $w_{min}$ for the given type of the transformers. If we used a stripline M7aM4bM8 as the transformer primary aligned over an equal-width stripline M6aM4bM8 secondary, the results are very similar: $w_{min} \approx$95 nm; at $w =$95 nm, $l_L =$1.58 μm, $l_q \approx$12.6 μm, and a 5% cross-talk is reached at $S_{min} =$1.65 μm. Hence, in the SFQ7ee process with Nb inductors, the number density of the AQFPs is limited only by the length of the inductors, especially by the inductor $L_q$, and given by (22a). It is shown in Fig. 14 by the bottom curves. In the practical range of the linewidth, $w \gtrsim$80 nm, the AQFP number density is below $4 \times 10^6$ cm$^{-2}$.

### B. Thin-film Kinetic Inductors

The only way to significantly increase $n_{AQFP}$, close to 18M per cm$^2$, and utilize advantages of striplines for lowing the

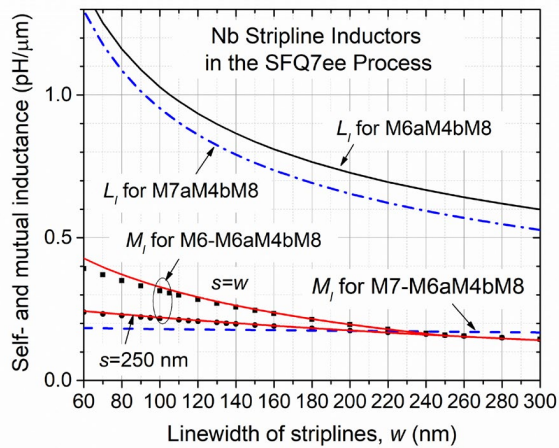

(a)

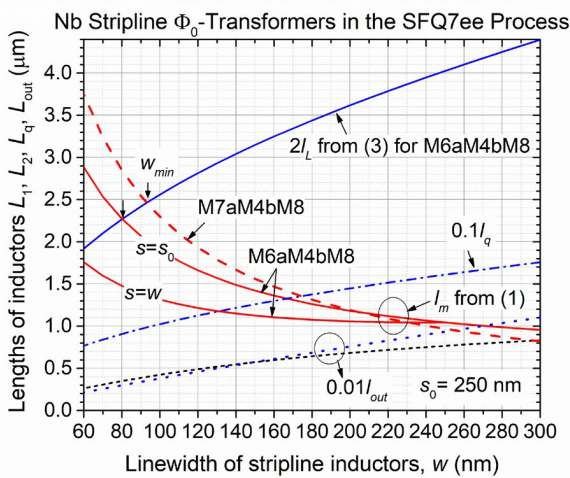

(b)

**Fig. 13.** (a) Self- and mutual inductance per unit length of niobium stripline inductors in the SFQ7ee process nodes with 10 superconducting layers and parameters: $H$ =1415 nm (dielectric thickness between M4 and M8 ground planes, $d_1$ =615 nm (dielectric thickness between M4 and M6), and $d_2$ =1015 nm (dielectric thickness between M4 and M7). Thickness of all the Nb layers is 200 nm.

(b) Lengths of stripline inductors in the AQFPs: solid blue line – length $l_L$ (3) of the AQFP cell inductor $L_2$ if made as M6aM4bM8 stripline; dash red curve – length $l_m$ (1) of the AQFP primary providing $\Phi_0$ flux excitation if made as M7aM4bM8 stripline; solid red curves - length $l_m$ (1) of the AQFP primary providing $\Phi_0$ flux excitation if made as M6aM4bM8 stripline spaced at either 250 nm from $L_2$ (top red curve) or at $s = w$ (bottom red curve). Dash dot blue cure is $0.1l_q$ (15), which equals $0.8l_L$ if inductor $L_q$ is the same M6aM4bM7 stripline as the $L_2$. Short black dash and blue dot curves at the bottom are $0.01l_{out}$ (18b), $l_{out}$ is the maximum length of the output inductor $L_{out}$ if it is made as M7aM4bM8 stripline (blue dot) or as M6aM4bM8 stripline spaced from $L_q$ at $s = 250$ nm (black short dash).

crosstalk, is to reduce $l_q$ down to $l_q \sim S_{min}$ by using a kinetic inductor, as shown in Fig. 14 by the two top curves. To achieve this, the process layer stack would need a layer of kinetic inductors near the JJ layer, in addition to geometrical inductors and ac power transmission lines formed by Nb layers.

The first option for adding a layer of kinetic inductors near the JJ layer is the following. This layer, labeled K5, can replace the layer of resistors R5 in the SFQ5ee process stack if shunt resistors for Josephson junctions are not needed or can be moved below the layer of JJs as in the SC1 and SC2 processes [41], [11].

With this modification, the contact to the junctions' top electrode, J5 and to the kinetic inductors K5 is still made by the layer M6 through vias, respectively, C5J and C5K, where label C5K replaces C5R in the existing SFQ5ee process. In case the shunt resistors are needed, they can be moved into the position R4 below the layer M5, as in the SC1 process [41].

If the K5 layer replaces a thin layer of resistors in the R5 position, only a thin film with similar thickness can be used, e.g., a 40-nm $Mo_2N$ film used in the SFQ5ee process for bias inductors on the L0 layer. It has the sheet inductance of 8 pH/sq [36]. Another option would be a thin, 40 nm to 50 nm, NbN film with a similar sheet inductance [42]. Using these high kinetic inductance materials would result in very short inductors having very small mutual coupling to other inductors. For instance, at $w_q = 0.25$ μm this would give $l_q \approx 0.33$ μm, which is too short a length to provide sufficient coupling to the output inductor $L_{out}$. Indeed, the strongest coupling to a K5 trace located at $d_1$ =505 nm ($h_1$ =525 nm) is achieved if the $L_{out}$ trace is on the neighboring layer M6 at $d_1$ =615 nm ($h_1$ =715 nm). Then, using (28) or numerical simulations, we get the linear mutual inductance $M_{ql} \approx 0.32$ pH/μm and the total mutual inductance $M_q = M_{ql}l_q \approx 0.1$ pH. According to (18b), this small mutual inductance would allow to transfer the output data only over distances below about 2 μm.

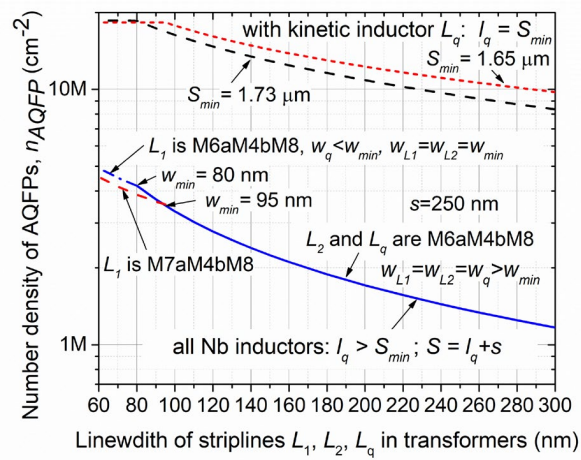

**Fig. 14.** Number density of AQFPs in the SFQ7ee process using Nb striplines M6aM4bM8 and M7aM4bM8 in the transformers (the bottom curves) and using a kinetic inductor for $L_q$ (the top two curves). Solid and dash-dot blue curves at the bottom correspond to all inductors made as M6aM4bM8 striplines spaced at $s$ =250 nm; below $w_{min}$ =80 nm only the width of the inductor $L_q$ can be reduced to decrease its length and increase the number density. The bottom red dash curve corresponds to the excitation transformer primary $L_1$ made as M7aM4bM8 stripline aligned over the equal-width secondary M6aM4bM7; below $w_{min}$ =95 nm only the width of the inductor $L_q$ can be reduced to decrease its length $l_q$ and increase the number density. The top red short dash curve corresponds to the ac excitation transformers with M7aM4bM8 primary, having lower cross talk ( $S_{min}$ =1.65 μm at $w = w_{min}$ ) than if the primary is a M6aM4bM8 stripline spaced at $s$ =250 nm from the secondary ($S_{min}$ =1.73 μm at $w = w_{min}$). Both top curves assume that length of the inductor $L_q$ is adjusted to the value of $S_{min}$ shown, the 5% cross talk spacing, by using a kinetic inductor with the required sheet inductance.

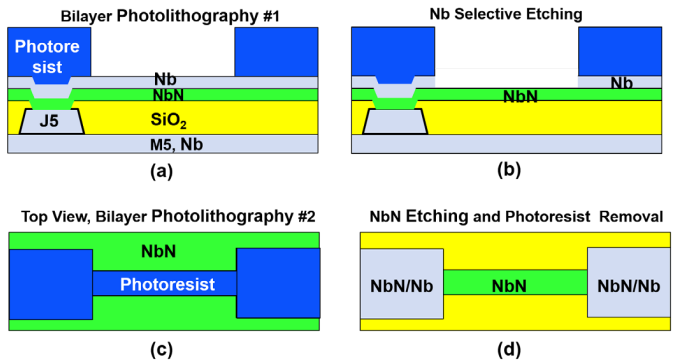

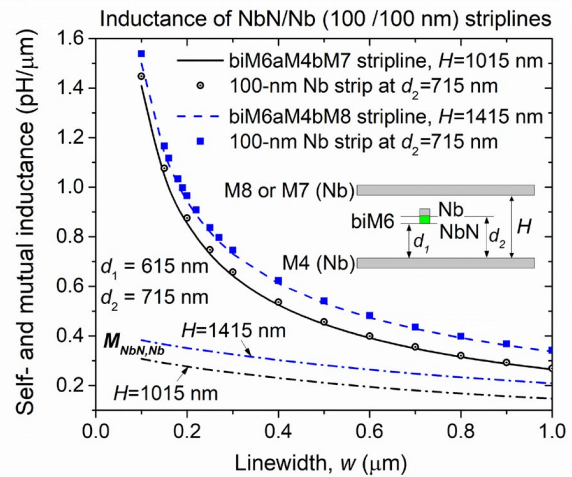

**Fig. 15.** Processing an in-situ deposited NbN/Nb bilayer, layer M6, to form kinetic inductors and geometrical inductors on the layer contacting Josephson junctions, J5: (a) cross section of the structure after the first photolithography; (b) after selective etching of the top Nb layer; (c) - top view of the second photolithography defining kinetic inductors; (d) – final top view of the patterned composite inductor after NbN etching and photoresist removal. This inductor consists of a geometrical inductance part formed by the full bilayer and the kinetic part formed by the patterned NbN.

**Fig. 16.** Inductance per unit length of NbN/Nb bilayer, $L_{bi}$ (thickness of the NbN and Nb layers is 100 nm) as layer M6 in striplines biM6aM4bM7 (solid black curve) and biM6aM4bM8 (blue dash curve) calculated using (32). Cross section of the structures is shown in the Inset. Thicknesses of all layers correspond to the SFQ7ee process: $t_{M4} = t_{biM6} = t_{M7} = t_{M8} = 200$ nm and dielectric thickness are given in the figure. Inductance of the 100-nm Nb strip of the bilayer, $L_{Nb}$ used in (32) is shown by open black and solid blue points. Mutual inductance per unit length between the layers of the bilayer, $M_{NbN,Nb}$ used in (32) is shown by the dash-dot curves for two dielectric thicknesses $H$ between the ground planes, corresponding to the two types of the striplines. All values were simulated using wxLC software.

Hence, for transferring the output data to larger distances we need to use smaller sheet inductances in the range from 1.5 to 1.75 pH/sq determined by the largest desired length $l_q \approx S_{min}$. Increasing thickness of the K5 layer significantly for achieving lower inductance is not desirable because this would require a full planarization of the K5 and increase the total dielectric thickness between the layers M5 and M6. The latter would make filling in of etched I5 vias by Nb of the M6 layer more difficult and reduce the via critical current. Therefore, we think that this approach is not practical and consider below another option.

### C. NbN/Nb Bilayer Kinetic/Geometric Inductors

The second option is to implement bilayer inductors – deposit the currently 200-nm M6 layer as a NbN/Nb bilayer. By patterning individual layers of this bilayer independently, we can create inductors in a very wide range of inductance values while maintaining an appropriate level of mutual inductance between them as explained below.

Inductance of a bilayer, $L_{bi}$ can be calculated as a parallel connection of inductances of the individual layers $L_{NbN}$ and $L_{Nb}$, assuming a sharp step-like change in the current density at the NbN/Nb interface

$$L_{bi} = \frac{L_{NbN}L_{Nb} - M_{NbN,Nb}^2}{L_{NbN} + L_{Nb} - 2M_{NbN,Nb}}, \quad (32)$$

where $M_{NbN,Nb}$ is the aiding mutual inductance between the NbN and Nb layers of the bilayer. Our measurements [42] show that magnetic field penetration depth (London penetration depth) in reactively sputtered NbN films is 510 nm, whereas it is 90 nm in the deposited Nb [12], [33]. Hence, in any practical range of the thickness of individual layers, $L_{NbN} \gg L_{Nb}$ due to a much larger kinetic inductance of the NbN film, while geometrical inductances of both layers are nearly the same due to their close geometry and location. Also, $L_{NbN} \gg M_{NbN,Nb}$ because the mutual inductance is smaller than the geometrical inductance of each layer, and $M_{NbN,Nb} < L_{Nb}$. As a result, inductance of the bilayer (per unit length) is completely determined by the inductance of the top Nb layer, $L_{bi} \lesssim L_{Nb}$.

Dependence of inductance of the equal thicknesses, 100 nm / 100 nm, Nb/NbN bilayer M6 (biM6) per unit length calculated using (32) as a function of the bilayer linewidth is shown in Fig. 16 for biM6aM4M7 and biM6aM4bM8 striplines along with the inductance of the top Nb layer of the bilayer. In the entire range of the linewidths, the bilayer stripline inductance is only about 2% lower than of the stripline using just the 100 nm thick top Nb layer of the bilayer, Hence, for all practical calculations $L_{bi} \approx L_{Nb}$.

A desired value $L$ of an inductor with length $l$, can be obtained by etching the top Nb layer from the bilayer over length $l_{NbN}$. Neglecting inductance associated with electrical current redistribution between the bilayer and the bottom NbN layer near the ends of the etched Nb, the resultant inductance can be treated as a serial connection of the NbN (mostly kinetic) inductor and the full bilayer (mostly geometric) inductor, giving

$$L = l_{NbN}L_{NbN} + (l - l_{NbN})L_{bi}. \quad (33)$$

As was shown in [33], mutual inductance between two inductors with small cross section does not depend on their superconducting properties. Therefore, in the first approximation, mutual inductance between the partially etched bilayer and Nb stripline inductor on layer M7 can be represented as

$$M = l_{NbN}M_{NbN,M7} + (l - l_{NbN})M_{biM6,M7}. \quad (34a)$$

and between the partially etched bilayer and the bilayer strip M6 as

$$M = l_{NbN}M_{NbN,biM6} + (l - l_{NbN})M_{biM6,biM6}, \quad (34b)$$

where $M_{NbN,biM6}$ and $M_{NbN,M7}$ are mutual inductances per unit length between, respectively, a stripline in the bottom NbN layer of the bilayer and the parallel stripline in the full bilayer, and between the NbN stripline and Nb stripline inductor M7; $M_{biM6,biM6}$ and $M_{biM6,M7}$ are mutual inductances per unit length between, respectively, two striplines made of the bilayer, and between the bilayer stripline and the M7 stripline. All these mutual inductances can be easily calculated using (31) because they do not depend on superconducting properties of the signal strips [33].

### D. AQFPs with NbN/Nb Bilayer Kinetic/Geometric Inductors

Let us estimate dimensions of an AQFP using Nb stripline M7aM4bM8 for the ac excitation transformer primary, NbN/Nb bilayer stripline biM6aM4bM8 for inductors $L_2$ (the secondary) and $L_{out}$, and a patterned bilayer for the inductor $L_q = 10.53$ pH. Mutual inductance per unit length of the bilayer stripline and M7aM4bM8 stripline, $M_{biM6,M7}$ is the same as of Nb striplines M6aM4bM8 and M7aM4bM8 shown in Fig. 13a because they have the same locations and geometrical dimensions; see (31) and [32]. Therefore, dependence of the required mutual length on the linewidth, $l_m(w)$ in (1) is also the same as shown in Fig.13b by the red dash curve. Using inductance per unit length of the 100/100 nm NbN/Nb bilayer, we calculate the total length of the excitation transfer secondary, $2l_L(w)$ shown in Fig. 17. The minimum linewidth of the primary determined from the condition $l_m(w)=2l_L(w)$ is $w_{min}=118$ nm, the largest value in all the considered cases due to the smallest value of $2l_L$ which in turn is a result of a higher inductance of the biM6 layer than of the 200-nm Nb M6 layer in the other cases. In the entire range of the linewidths, $2l_L + s \gg S_q = S_{min}$ for the stripline transformers; see Fig. 11. So, the crosstalk in the same row of AQFPs can be ignored.

The length $l_q$ can be adjusted to any desired value between the shortest $L_q/L_{NbN}(w)$ and the longest $L_q/L_{biM6}(w)$ values shown in Fig. 17 (respectively, short-dash magenta and dash-dot black lines) by fully or partially etching the top Nb layer of the bilayer; here $L_{NbN}(w)$ and $L_{biM6}(w)$ are width-dependent linear inductances of, respectively, the NbN layer and the NbN/Nb bilayer. To minimize crosstalk between the adjacent rows of AQFPs, we need $l_q + s \le S_{min}$. For the vertical stripline transformers, $S_{min}$ weekly depends on $w$ and $S_{min} = 1.65$ μm at a practical minimum width of the bilayer inductors $w = 100$ nm; see Fig. 11 (blue dash-dot and red short dash curves). The required etch length of Nb, i.e., the length of the kinetic part, $l_{NbN}$, of the inductor with $L_q=10.53$ pH and $l_q = 1.40$ μm (at $s = 250$ nm) and is shown in Fig. 17. This etch length is certainly within the capabilities of the existing SCE fabrication technology. Making shorter $l_q$ by etching a longer length of Nb off the bilayer strip is certainly possible. This would decrease the cell size but increase the crosstalk and also decrease the data output length $l_{out}$. In practice, the actual length will be a design-dependent and adjustable parameter.

Fig. 18 show the theoretical number density (29a)–(29b) of the AQFPs using the proposed fabrication process with bilayer inductors. As can be seen, densities above 22M cm$^{-2}$, corresponding to over 7M MAJ3 gates per cm$^2$, can be reached at very modest linewidth of all the inductors, which is fully within capabilities of the existing fabrication technology. This is the main result of this section.

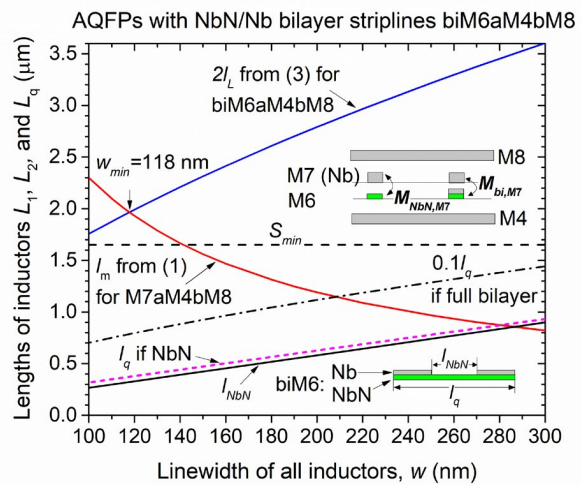

**Fig. 17.** Parameters of the ac excitation transformer and the AQFP cell inductors using Nb striplines M7aM4bM8 as the transformer primary and NbN/Nb bilayer striplines biM6aM4bM8 as the transformer secondary and the inductor $L_q$. The assumed thicknesses of the NbN and Nb layers is 100 nm. Below $w_{min}=118$ nm, the excitation current in the primary need to exceed the critical current of the Nb wire of 6 mA in order to induce a total flux of $\Phi_0$. The short-dash magenta line shows the length $l_q$ of the inductor $L_q=10.53$ pH if the inductor is completely made of the NbN film of the bilayer. The dash-dot black line shows $0.1l_q$ if the inductor is made of the bilayer. The solid blue line shows the length of the etched off Nb of the bilayer (see the bottom Inset), i.e., the length $l_{NbN}$ of the kinetic part of the inductor required to make $l_q = 1.4$ μm so that $l_q + s = S_{min} =1.65$ μm.

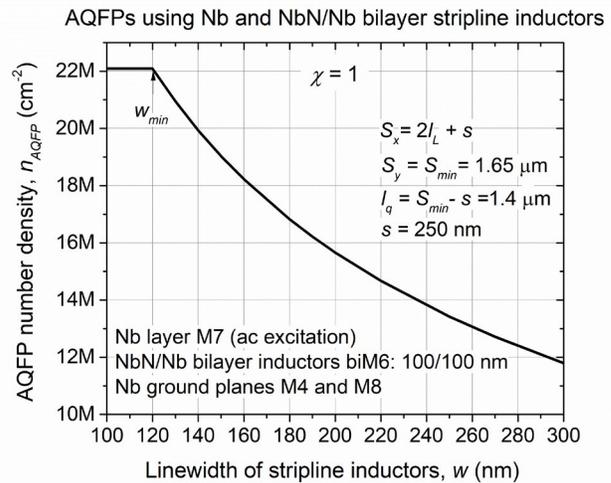

**Fig. 18.** Theoretical number density of AQFPs using NbN/Nb bilayer and Nb stripline inductors in the proposed advanced version of the SFQ7ee process; see text. For these AQFPs, the placement pith in the $x$-direction, $S_x$ is set by the physical width of the AQFP cell $2l_L$ and the minimum spacing $s$ used in the process, because it is larger than the crosstalk length $S_q = S_{min} =1.65$ μm. The $y$-direction placement pitch $S_y$ of the AQFP rows is set equal the 5% crosstalk spacing $S_{min} =1.65$ μm. This set the length $l_q$ of the AQFP inductor $L_q$ which is made by etching a part of Nb from the NbN/Nb bilayer.

## VI. Conclusion

We have considered two main factors limiting the integration scale (device number density) of superconductor integrated circuits using ac power for logic and/or memory cell excitation and clocking: critical current of superconducting transformers and their cross coupling. For each type of superconducting transformers, there is a minimum cross-section area $A_{min} \sim 0.013 \ \mu m^2$ and the corresponding minimum linewidth $w_{min} \sim 100$ nm of the ac power transmission line (the transformer primary) below which the superconducting current required to provide the flux excitation needed for the cell operation exceeds the critical current of the wire. This critical current as well as the mutual inductance set the minimum mutual coupling length between the transformer primary and secondary and hence the minimum size of the logic or memory cell in the $x$-direction (along the ac power line). On the other hand, reduction of the linewidth of the transformer secondary increases its kinetic inductance and, hence, decreases its length since the total inductance is set by the cell design parameter $\beta_L$. This length reduction decreases the mutual coupling in the transformer and prevents transformer miniaturization.

Mutual coupling (crosstalk) between adjacent transformer is strong and long-ranged if transformers use microstrip inductors. This may present a serious problem and also limit the scale of integration. Crosstalk diminishes exponentially with spacing between the transformer if stripline inductors are used. However, mutual coupling of striplines is much smaller than of the microstrip, which increases the size of the transformers.

Using parameters of AQFP cells as a typical example, we have estimated the maximum number density of AQFP circuits for various types of microstrip and stripline inductors which can be formed near the Josephson junctions in fully planarized fabrication processes for superconductor electronics developed at MIT Lincoln Laboratory. We have shown that, at the current SFQ5ee process minimum linewidth and spacing $w = s = 250$ nm, the theoretical AQFP number density is about 1M per cm². Reduction of Nb linewidth to about 60 nm in the future processes may increase the AQFP number density to a few million per cm².

We have shown that the circuit density can be substantially increased by using kinetic inductors, e.g., patterned NbN films instead of Nb, mainly geometrical, inductors. Since short strips of kinetic inductors have very small mutual coupling, we have proposed to use bilayer inductors, e.g., NbN/Nb bilayers consisting of a layer of kinetic inductor (material with large $\lambda$) covered by a layer of geometrical inductor (material with small $\lambda$, niobium). Partial patterning of the top Nb layer of the bilayer enables making kinetic inductors with a wide range of inductance values from the patterned bottom layer, whereas using the full bilayer allows for making small-value inductors and preserves sufficient mutual coupling.

Additional design flexibility is provided by selecting thicknesses of the individual layers in the bilayer. As an example, we have considered a 100 nm / 100 nm NbN/Nb bilayer as the layer M6 in a future fabrication process node SFQ7ee having nine superconducting layers, three of which are above the layer of Josephson junctions. We have calculated parameters of the transformers and dimensions of the AQFP cells using Nb striplines M7aMA4M8 for ac power delivery and striplines biM6aMabM8 with patterned Nb of the bilayer for the AQFP cell inductors. We have found that the proposed advanced fabrication process with bilayer inductors allows for the highest AQFP device number density among all the considered processes and options, reaching above 22M AQFPs per cm², corresponding to about 7M MAJ3 logic gates per cm², at modest linewidths $w \gtrsim 120$ nm which are within the capabilities of the fabrication equipment and existing fabrication processes for superconductor electronics.

The current status of the SFQ7ee process development and progress towards NbN/Nb bilayer inductors will be reported at the Applied Superconductivity Conference, ASC 2022 and published elsewhere.

Although all the calculations were done for the typical parameters of AQFP cells, the same conclusions and scaling estimates, with small modifications, are applicable to all other circuits using ac power and/or transformers such as RQL, nSQUID circuits [42], ac-biased SFQ circuits with ac-dc converters [43], and circuits with single flux quantum biasing [44], and likely to neuromorphic bioSFQ circuits [45], using flux transformers in artificial superconducting neurons.

*The estimated maximum number density of ac-powered (AQFP) circuits of about 22M cm⁻² should be considered as the upper limit to the achievable scale of integration of superconductor electronics utilizing ac excitation and clocking.* A separate discussion is required of potential applications of superconductor ac-clocked logics and memory for which this integration scale would be sufficient. For instance, whether it is sufficient for applications in general purpose or high-performance computing, data centers, artificial neural networks and neuromorphic processors, cold processors for quantum computers, cold processors for large arrays of cold sensors, etc.

## Acknowledgment

I am grateful to Vasili Semenov and Timur Filippov for many interesting discussions of scalability of superconductor electronics, to Vladimir Bolkhovsky for the numerous discussions of NbN films and bilayer inductors, and to Mark Gouker and Leonard Johnson for their interest in this work. This research was based upon work supported by the Under Secretary of Defense for Research and Engineering via Air Force Contract No. FA8702-15-D-0001. Any opinions, findings, conclusions or recommendations expressed in this material are those of the authors and do not necessarily reflect the views of the Under Secretary of Defense for Research and Engineering or the U.S. Government. Notwithstanding any copyright notice, U.S. Government rights in this work are defined by DFARS 252.227-7013 or DFARS 252.227-7014 as detailed above. Use of this work other than as specifically authorized by the U.S. Government may violate any copyrights that exist in this work. The U.S. Government is authorized to reproduce and distribute reprints for Governmental purposes notwithstanding any copyright annotation thereon.

## REFERENCES

[1] K.K. Likharev and V.K. Semenov, "RSFQ logic/memory family: a new Josephson-junction technology for sub-terahertz-clock-frequency digital systems," *IEEE Trans. Appl. Supercond.*, vol. 1, no. 1, pp. 3-28, Mar. 1991.

[2] W. Chen, A.V. Rylyakov, V. Patel, J.E. Lukens, and K.K. Likharev, "Superconducting digital frequency dividers operating up to 750 GHz," *Appl. Phys. Lett.*, vol. 73, pp. 2817-2819, Nov. 1998.

[3] K.K. Likharev, "Dynamics of some single flux quantum devices: I. Parametric Quantron," *IEEE Trans. Mag.*, vol. MAG-13, no. 1, pp. 242-244, Jan. 1977.

[4] V.K. Semenov, G.V. Danilov, and D.V. Averin, "Negative-inductance SQUID as the basic element of reversible Josephson-junction circuits," *IEEE Trans. Appl. Supercond.*, vol. 13, no. 2, pp. 938-943, June 2003.

[5] J. Ren and V.K. Semenov, "Progress with physically and logically reversible superconducting digital circuits," *IEEE Trans. Appl. Supercond.*, vol. 21, no. 3, pp. 780-786, June 2011.

[6] N. Takeuchi, Y. Yamanashi, and N. Yoshikawa, "Measurements of 10 zJ energy dissipation of adiabatic quantum-flux-parametron logic using a superconducting resonator," *Appl. Phys. Lett.*, vol. 102, Feb. 2013, Art. no. 052602.

[7] V.K. Semenov, Y.A. Polyakov, S.K. Tolpygo, "AC-biased shift registers as fabrication process benchmark circuits and flux trapping diagnostic tool," *IEEE Trans. Appl. Supercond.*, vol. 27, no. 4, June 2017, Art. no. 1301409.

[8] *500-qubit Advantage chip, D-Wave Systems, Inc.* On-line. Available: https://www.dwavesys.com/solutions-and-products/systems/

[9] *Transistor count*, On-line. Available: https://en.wikipedia.org/wiki/Transistor_count

[10] S.K. Tolpygo, V. Bolkhovsky, R. Rastogi *et al.*, "A 150-nm process node of an eight-Nb-layer fully planarized process for superconductor electronics," *IEEE CSC & ESAS Superconductivity News Forum (global edition)*, no. 49, Mar. 2021. Invited presentation Wk1EOr3B-01 at Appl. Supercond. Conf., ASC 2020, Oct. 29, 2020. [Online] Available: https://snf.ieeecsc.org/sites/ieeecsc.org/files/documents/snf/abstracts/STP669%20Tolpygo%20invited%20pres.pdf

[11] S.K. Tolpygo, E.B. Golden, T.J. Weir, V. Bolkhovsky, "Inductance of superconductor integrated circuit features with sizes down to 120 nm," *Supercond. Sci. Technol.*, vol. 34, June 2021, Art. no. 085005, doi: 10.1088/1361-6668/ac04b9

[12] S.K. Tolpygo and V.K. Semenov, "Increasing integration scale of superconductor electronics beyond one million Josephson junctions," *J. Phys.: Conf. Ser.*, vol. 1559, Art. no. 012002, 2020.

[13] V.K. Semenov, Yu. A. Polyakov, and S.K. Tolpygo, "Very large scale integration of Josephson-junction-based superconductor random access memories," *IEEE Trans. Appl. Supercond.*, vol. 29, no. 5, Aug. 2019, Art. no. 1302809.

[14] *EMD4E001G-1Gb Spin-transfer Torque MRAM.* On-line. Available: https://www.everspin.com/family/emd4e001g?npath=3557

[15] S.K. Tolpygo, "Superconductor electronics: scalability and energy efficiency issues," *Low Temp. Phys. / Fizika Nizkikh Temperatur*, vol. 42, no. 5, pp. 463-485, May 2012; doi: 10.1063/1.4948618

[16] K.K. Likharev, G.M. Lapir, and V.K. Semenov, "Properties of the superconducting loop closed with the Josephson junction with variable critical current", *Sov. Techn. Phys. Lett.*, vol. 2, pp. 809-814, Sov. 1976.

[17] K.K. Likharev, S.V. Rylov, and V.K. Semenov, "Reversible coveyer computation in array of parametric quantrons," *IEEE Trans. Mag.*, vol. MAG-21, no. 2, pp. 947-950, Mar. 1985.

[18] K. Loe and E. Goto, "Analysis of flux input and output Josephson pair device," *IEEE Trans. Mag.*, vol. MAG-21, no. 2, pp. 884-887, Mar. 1985, doi: 10.1109/TMAG.1985.1063734.

[19] E. Goto and K.F. Loe, "*DC Flux Parametron. A New Approach to Josephson Junction Logic*," Singapure: World Scientific, 1986.

[20] Y. Harada, H. Nakane, N. Miyamoto *et al.*, "Basic operation of quantum flux parametron," *IEEE Trans. Magn.*, vol. MAG-23, no. 5, pp. 3801-3807, Sep. 1987, doi: 10.1109/TMAG.1987.106557.

[21] M. Hosoya, W. Hioe, J. Casas, "Quantum flux parametron: a single quantum flux device for Josephson supercomputer," *IEEE Trans. Appl. Supercond.*, vol. 1, no. 2, pp. 77–89, Jun. 1991.

[22] N. Takeuchi, D. Ozawa, Y. Tamanashi, and N. Yoshikawa, "An adiabatic quantum flux parametron as an ultra-low-power logic device," *Supercond. Sci. Technol.*, vol. 26, no. 3, Mar. 2013, Art. no. 035010.

[23] K. Inoue, N. Takeuchi, K. Ehara *et al.*, "Simulation and experimental demonstration of logic circuits using an ultra-low-power adiabatic quantum-flux-parametron," *IEEE Trans. Appl. Supercond.*, vol. 23, no. 3, Jun. 2013, Art. no. 1301105.

[24] V.K. Semenov, G.V. Danilov, and D.V. Averin, "Classical and quantum operation modes of the reversible Josephson-junction logic circuits," *IEEE Trans. Appl. Supercond.*, vol. 17, no. 2, pp. 455–461, Jun. 2007.

[25] Q.P. Herr, A.Y. Herr, O.T. Oberg, and A.G. Ioannidis, "Ultra-low-power superconducting logic," *J. Appl. Phys.*, vol. 109, 2011, Art. ID 103903.

[26] A. Y. Herr, Q.P. Herr, O.T. Oberg, O. Naaman, J.X. Przybysz, P. Borodulin, and S.B. Shauck, "An 8-bit carry look-ahead adder with 150 ps latency and submicrowatt power dissipation at 10 GHz," *J. Appl. Phys.*, vol. 113, 033911, Jan. 2013.

[27] Q.P. Herr, J. Osborne, M.J.A. Stoutimore, H. Hearne, R. Selig, J. Vogel, E. Min, V.V. Talanov, and A.Y. Herr, "Reproducible operating margins on a 72800-device digital superocnudcting chip," *Supercond. Sci. Technol.*, vol. 28, 124003, Oct. 2015.

[28] L. Ginzburg and L.D. Landau, "On the theory of superconductivity," *Zh. Eksp. Teor. Fiz.*, vol. 20, p. 1064, 1950. In Russian. For English version, see, e.g., V.L. Ginzburg "*On Superconductivity and Superfluidity: A Scientific Autobiography*," Germany, Berlin: Springer-Verlag, 2009, pp. 112-135.

[29] J. Bardeen, L.N. Cooper, and J.R. Schrieffer, "Theory of superconductivity," *Phys. Rev.*, vol. 108, no. 5, pp. 1175-1204, Dec. 1957.

[30] J. Bardin, "Critical fields and currents in superconductors," *Rev. Mod. Phys.*, vol. 34, no. 4, pp. 667–681, Oct. 1962.

[31] S.K. Tolpygo, V. Bolkhovsky, T. Weir, L. Johnson, W.D. Oliver, and M.A. Gouker, "Deep sub-micron stud-via technology of superconductor VLSI circuits," *Supercond. Sci. Technol.*, vol. 27, 025016, Jan. 2014, doi:10.1088/0953-2048/27/2/025016.

[32] S.K. Tolpygo, E.B. Golden, T.J. Weir, and V. Bolkhovsky, "Mutual and self-inductance in planarized multilayered superconductor integrated circuits: Microstrips, striplines, bends, meanders, ground plane perforations," *IEEE Trans. Appl. Supercond.*, vol. 32, no. 5, pp. 1-31, Aug. 2022, Art. no. 1400331, doi: 10.1109/TASC.2022.3162758

[33] N. Takeuchi, S. Nagasawa, F. China, T. Ando, M. Hidaka, Y. Yamanashi, and N. Yoshikawa, Adiabatic quantum-flux-parametron cell library designed using a 10 kA cm⁻² niobium fabrication process," *Supercond. Sci. Technol.*, vol. 30, Jan 2017, Art. no. 035002.

[34] S. K. Tolpygo, V. Bolkhovsky, T. J. Weir *et al.*, "Advanced fabrication processes for superconducting very large scale integrated circuits," *IEEE Trans. Appl. Supercond.*, vol. 26, no. 3, Apr. 2016, Art. no. 1100110

[35] S.K. Tolpygo, V. Bolkhovsky, D.E. Oates, R. Rastofi, A.L. Day, T.J. Weir, A. Wynn, and L.M. Johnson, "Superconductor electronics fabrication process with MoNₓ kinetic inductors and self-shunted Josephson junctions," *IEEE Trans. Appl. Supercond.*, vol. 28, no.4, Jun. 2018, Art. no. 1100212.

[36] N. Takeuchi, Y. Yamanashi, and N. Yoshikawa, "Adiabatic quantum-flux-parametron cell library adopting minimalist design, " *J. Appl. Phys.*, vol. 117, Art. no. 173912, May 2015, doi: 10.1063/1.4919838.

[37] N. Takeuchi, K. Arai, and N. Yoshikawa, "Directly coupled adiabatic superconductor logic," *Supercond. Sci. Technol.*, vol. 33, no. 6, Art. no. 065002, May 2020, doi: 10.1088/1361-6668/ab87ad.

[38] M.M. Khapaev, "Extraction of inductances of a multi-superconductor transmission line," *Supercond. Sci. Technol.*, vol. 9, pp. 729-733, 1996.

[39] D.E. Kirichenko, S. Sarwana, and A.F. Kirichenko, "Zero static power dissipation biasing of RSFQ circuits," *IEEE Trans. Appl. Supercond.*, vol. 21, no. 3, pp. 776-779, Jun. 2011, doi: 10.1109/TASC.2010.2098432

[40] S. K. Tolpygo *et al.*, "Advanced fabrication processes for superconductor electronics: Current status and new developments," *IEEE Trans. Appl. Supercond.*, vol. 29, no. 5, pp. 1-13, Aug. 2019, Art no. 1102513, doi: 10.1109/TASC.2019.2904919.

[41] S.K. Tolpygo, E.B. Golden, T. Weir, and V. Bolkhovsky, unpublished

[42] V. K. Semenov, G. V. Danilov and D. V. Averin, "Negative-inductance SQUID as the basic element of reversible Josephson-junction circuits," *IEEE Trans. Appl. Supercond.*, vol. 13, no. 2, pp. 938-943, June 2003, doi: 10.1109/TASC.2003.814155.

[43] V. K. Semenov, Y. A. Polyakov and S. K. Tolpygo, "New AC-powered SFQ digital circuits," *IEEE Trans. on Appl. Supercond.*, vol. 25, no. 3, pp. 1-7, June 2015, Art no. 1301507, doi: 10.1109/TASC.2014.2382665.

[44] V. K. Semenov, E. B. Golden and S. K. Tolpygo, "SFQ bias for SFQ digital circuits," *IEEE Trans. Appl. Supercond.*, vol. 31, no. 5, pp. 1-7, Aug. 2021, Art no. 1302207, doi: 10.1109/TASC.2021.3067231.

[45] V. K. Semenov, E. B. Golden and S. K. Tolpygo, "A new family of bioSFQ logic/memory cells," *IEEE Trans. Appl. Supercond.*, vol. 32, no. 4, pp. 1-5, June 2022, Art no. 1400105, doi: 10.1109/TASC.2021.3138369.